%% file: main.tex
\documentclass[letterpaper,twocolumn,10pt]{article}
\usepackage{usenix-2020-09}

\usepackage{tikz}
\usepackage{amsmath}

\if11
\else
\fi

\usepackage{xurl}
\usepackage{tikz}
\usepackage{amsmath}

\usepackage{etoolbox}

\include{cmds}

\usepackage[utf8]{inputenc}
\usepackage[T1]{fontenc} % Use 8-bit encoding that has 256 glyphs

\usepackage[english]{babel} % Language hyphenation and typographical rules

\let\labelindent\relax % incompatibility with IEETran - https://tex.stackexchange.com/questions/170772/command-labelindent-already-defined
\usepackage[inline]{enumitem} %% gives an error if uncommented

\usepackage{xspace}

\newcommand{\mypara}[1]{\vspace{5pt}\noindent{\textbf{{#1}}\xspace}}

\usepackage{graphicx}
\usepackage{blindtext} 

\usepackage{makecell} %for multile line break in cells

\usepackage{amssymb} %For ticks

\usepackage{dashrule}

\usepackage{listings}
\usepackage{array} % For bugs summary table

\usepackage{inconsolata} % to have nice mono fonts

\usepackage{array}
\usepackage{tabularx}
\usepackage{booktabs}
\usepackage{arydshln}

\usepackage{multirow}

\newcolumntype{L}[1]{>{\raggedright\let\newline\\\arraybackslash\hspace{0pt}}m{#1}}
\newcolumntype{C}[1]{>{\centering\let\newline\\\arraybackslash\hspace{0pt}}m{#1}}
\newcolumntype{R}[1]{>{\raggedleft\let\newline\\\arraybackslash\hspace{0pt}}m{#1}}

\usepackage{soul}

\usepackage{wasysym}
\usepackage{amssymb}% http://ctan.org/pkg/amssymb
\usepackage{pifont}% http://ctan.org/pkg/pifont
\newcommand{\xmark}{\ding{55}}%
\usepackage{lscape}
\newcommand{\yes}{\CIRCLE}
\newcommand{\no}{\Circle}
\newcommand{\half}{\LEFTcircle}

\if10
\usepackage[]{graphicx}
\usepackage{scalerel}
\def\good{\scalerel*{\includegraphics{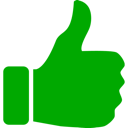}}{O}}
\def\bad{\scalerel*{\includegraphics{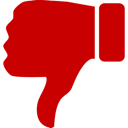}}{g}}
\else
\newcommand{\good}{\yes}
\newcommand{\bad}{\no}
\fi

\usepackage{makecell}

\newcommand{\project}{\textsc{ReZone}\xspace}

\newcommand\cveshort[1]{\href{https://www.cvedetails.com/cve/CVE-#1}{#1}}
\newcommand\cve[1]{\href{https://www.cvedetails.com/cve/CVE-#1}{CVE-#1}}
\usepackage{tikz}

\begin{document}
%-------------------------------------------------------------------------------

%don't want date printed
\date{}

% make title bold and 14 pt font (Latex default is non-bold, 16 pt)
\title{\Large \bf \project: Disarming TrustZone with TEE Privilege Reduction}

\author{%
{\rm David Cerdeira$^\dag$, José Martins$^\dag$, Nuno Santos$^\ddag$, Sandro Pinto$^\dag$}\\ % \\[1ex] % Your name
{\small $^\dag$Centro ALGORITMI, Universidade do Minho, $^\ddag$INESC-ID / Instituto Superior Técnico, Universidade de Lisboa } \\ % Your institution
{\small	\{david.cerdeira, jose.martins, sandro.pinto\}@dei.uminho.pt, nuno.m.santos@tecnico.ulisboa.pt} \\ % Your email address
% copy the following lines to add more authors
% \and
% {\rm Name}\\
%Name Institution
} % end author
% LATEX dagger

\maketitle

%-------------------------------------------------------------------------------
\begin{abstract}
%-------------------------------------------------------------------------------
In TrustZone-assisted TEEs, the trusted OS has unrestricted access to both secure and normal world memory. Unfortunately, this architectural limitation has opened an aisle of exploration for attackers, which have demonstrated how to leverage a chain of exploits to hijack the trusted OS and gain full control of the system, targeting (i) the rich execution environment (REE), (ii) all trusted applications (TAs), and (iii) the secure monitor. In this paper, we propose \project. The main novelty behind \project design relies on leveraging TrustZone-agnostic hardware primitives available on commercially off-the-shelf (COTS) platforms to restrict the privileges of the trusted OS. With \project, a monolithic TEE is restructured and partitioned into multiple sandboxed domains named \textit{zones}, which have only access to private resources. We have fully implemented \project for the i.MX 8MQuad EVK and integrated it with Android OS and OP-TEE. We extensively evaluated \project using microbenchmarks and real-world applications. \project can sustain popular applications like DRM-protected video encoding with acceptable performance overheads. We have surveyed 80 CVE vulnerability reports and estimate that \project could mitigate 86.84\% of them.

\end{abstract}

\pagestyle{empty} %% no page numbers

% Sections ------------------------------------------------------------------------

\input{main_body.tex}

%%%%%%%%%%%%%%%%%%%%%%%%%%%%%%%%%%%%%%%%%%%%%%%%%%%%%%%%%%%%%%%%%%%%%%%%%%%%%%%%
\end{document}

%% file: cmds.tex
\newcounter{vnum}
\stepcounter{vnum}
\renewcommand*{\thevnum}{%
    V
    \ifnum\value{vnum}<10 0\fi
    \arabic{vnum}%
}

\newcounter{fnum}
\stepcounter{fnum}
\renewcommand*{\thefnum}{%
    L%
    \ifnum\value{fnum}<10 0\fi
    \arabic{fnum}%
}

\newcounter{inum}
\stepcounter{inum}
\renewcommand*{\theinum}{%
    I%
    \ifnum\value{inum}<10 0\fi
    \arabic{inum}%
}

\newcounter{dnum}
\stepcounter{dnum}
\renewcommand*{\thednum}{%
    D%
    \ifnum\value{dnum}<10 0\fi
    \arabic{dnum}%
}

\iftrue
\fi

%% file: main_body.tex
%=======================================================================
%> [INTRO]
%=======================================================================	
\section{Introduction}

Arm TrustZone~\cite{Pinto2019, Pinto2019_2} is a technology embedded into Arm processors shipped in billions of mobile phones and embedded devices. Vendors and Original Equipment Manufacturers (OEM) rely on TrustZone for deploying Trusted Execution Environments (TEEs), which protect the execution of sensitive programs named Trusted Applications (TAs). Some TAs implement kernel-level services of the operating system (OS), e.g., for user authentication or file disk encryption~\cite{androidwhitepaper}. Other TAs provide shared user-level functionality, e.g., DRM media decoders~\cite{TZVideoPath} or online banking services~\cite{trustonicbanking}. TEEs themselves consist of a trusted software stack offering API and runtime support for hosting TAs. TEEs such as Qualcomm's QSEE and OP-TEE protect the confidentiality and integrity of TAs' memory state, thereby ensuring it cannot be inspected or tampered with by a potentially compromised OS.

To protect the TAs, TEEs leverage mechanisms provided by TrustZone. In Armv7-A/Armv8-A, this technology provides two execution environments named \textit{normal world} and \textit{secure world}\footnote{As per Arm's recent documentation~\cite{tzarmv8a}, \textit{secure world} can be referred to as \textit{trusted world}; we adopt the previous and more familiar terminology~\cite{Pinto2019}.}. The normal world runs a rich software stack, named Rich Execution Environment (REE), comprised of full-blown OS and applications; it is generally considered untrusted. The secure world runs a smaller software stack consisting of TEE and TAs. TrustZone enforces system-wide isolation between worlds and provides controlled entry points for world switching whenever the REE invokes TAs' services.

\begin{figure}[t]
    \centering
    \includegraphics[width=0.98\linewidth]{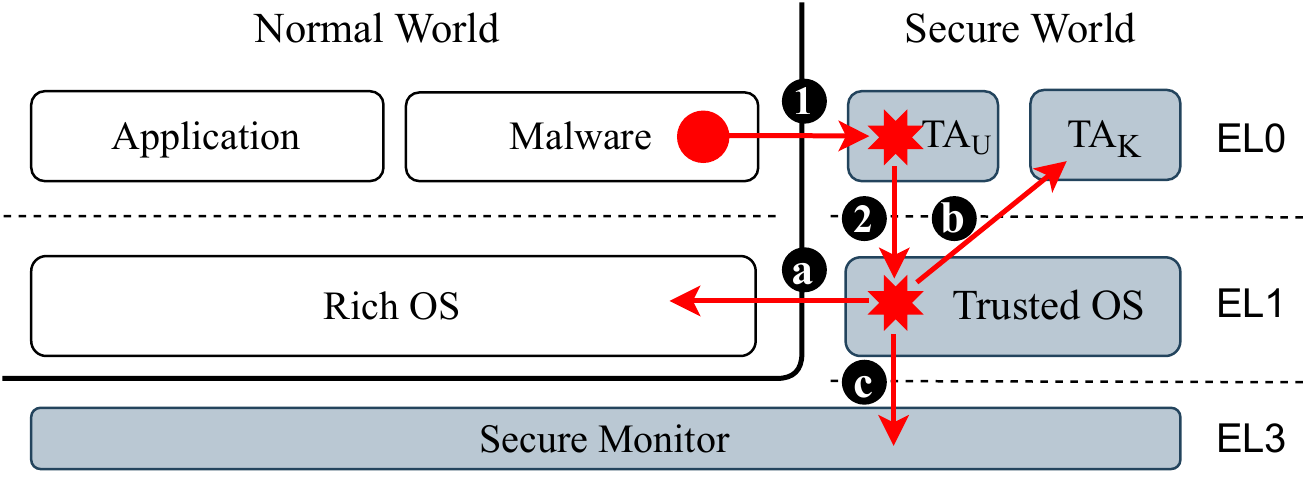}
    \vspace{-0.3cm}
    \caption{Privilege escalation attack in a TrustZone-assisted TEE: (1) hijack a user-level TA, e.g., exploiting vulnerability in the Widevine TA (CVE-2015-6639)~\cite{gb-exploit4}, (2) hijack the trusted OS, e.g., exploiting vulnerability in syscall interface of Qualcomm's QSEOS (CVE-2016-2431)~\cite{gb-exploit5}. Once in control of the trusted OS, other attacks can be launched: (a) control the rich OS, e.g., backdoor in the Linux kernel~\cite{gb-exploit6}, (b) control a kernel-level TA, e.g., to extract full disk encryption keys of Android's KeyMaster service~\cite{gb-keymaster}, or (c) control the secure monitor, e.g., to unbrick the device bootloader~\cite{gb-bootloader}.}
    \label{fig:intro_trustzone}
    \vspace{-0.2cm}
\end{figure}

However, in the TrustZone security architecture, the \textit{trusted OS} -- i.e., one of TEE's core components -- runs with an excess of privileges, exposing the entire system to catastrophic privilege escalation attacks if this component gets compromised. Figure~\ref{fig:intro_trustzone} illustrates these attacks. It shows the TEE stack, consisting of secure monitor and trusted OS. The secure monitor runs in EL3 -- the most privileged exception level -- and provides essential mechanisms for world switching and platform management. The trusted OS runs in S.EL1 -- i.e., secure kernel mode -- and implements memory management, scheduling, IPC, and common services for TAs. TAs run in unprivileged S.EL0 -- i.e., secure user mode. TAs are directly exposed to malicious requests coming from normal world applications. If a TA's control flow gets hijacked as a result of exploiting a software vulnerability (1), the attacker can control that TA. If the attacker can further hijack the trusted OS by exploiting another vulnerability (2), he can gain full control of the system, including the REE in the normal world (a), all the TAs (b), and the secure monitor (c).

This level of control is possible due to a fundamental violation of the principle of least privilege where an excess of privileges is dangerously (and unnecessarily) granted to some protection modes in the secure world. In particular, the attacks just described leverage the fact that S.EL1 gives full permission to access and reconfigure both secure and non-secure memory regions. Such level of control, however, should be reserved only to the highest protection domain, i.e., EL3. Similar problems mostly carryover to Armv8.4-A and onward, with the new S.EL2 hypervisor mode for the secure world having the same privileges over the monitor and the normal world as S.EL1 had up until Armv8.4-A; if the secure hypervisor is exploited the entire system will be compromised. Armv9-A is also susceptible to some of these concerns, for instance, allowing S.EL2 to arbitrarily control the normal world state.

Unfortunately, numerous devices face serious risks of being compromised by attacks of this nature. In the past, security researchers have demonstrated how these attacks can be mounted by targeting the trusted OS through a chain of vulnerability exploits as described in the caption of Figure~\ref{fig:intro_trustzone}. Notably, recent studies~\cite{Cerdeira2020} have shown that commercial TEE systems have plenty and serious vulnerabilities, most of them affecting user-level TAs and trusted OS. This suggests that the current attack surface for TEE systems is seriously enlarged.

Despite the abundant research on TrustZone security~\cite{Pinto2019,Cerdeira2020}, there is a lack of practical system defenses if the trusted OS is hijacked. TEE systems like Sanctuary~\cite{Brasser2019} aim to reduce the attack surface of the trusted OS by relocating TAs onto user-level enclaves in the normal world. However, due to portability reasons, vendors and OEMs continue deploying TAs in commercial TEEs, therefore exposing many devices to potential attacks. Another approach is to confine the TEE stack inside a secure world sandbox, e.g., by leveraging same privilege isolation~\cite{Kwon19,Li2019} or hardware virtualization~\cite{hafnium}. However, these approaches do not solve the fundamental problem of the excess of privileges granted to S.EL1 (or S.EL2).

\begin{figure}[t]
    \centering
    \includegraphics[width=0.99\linewidth]{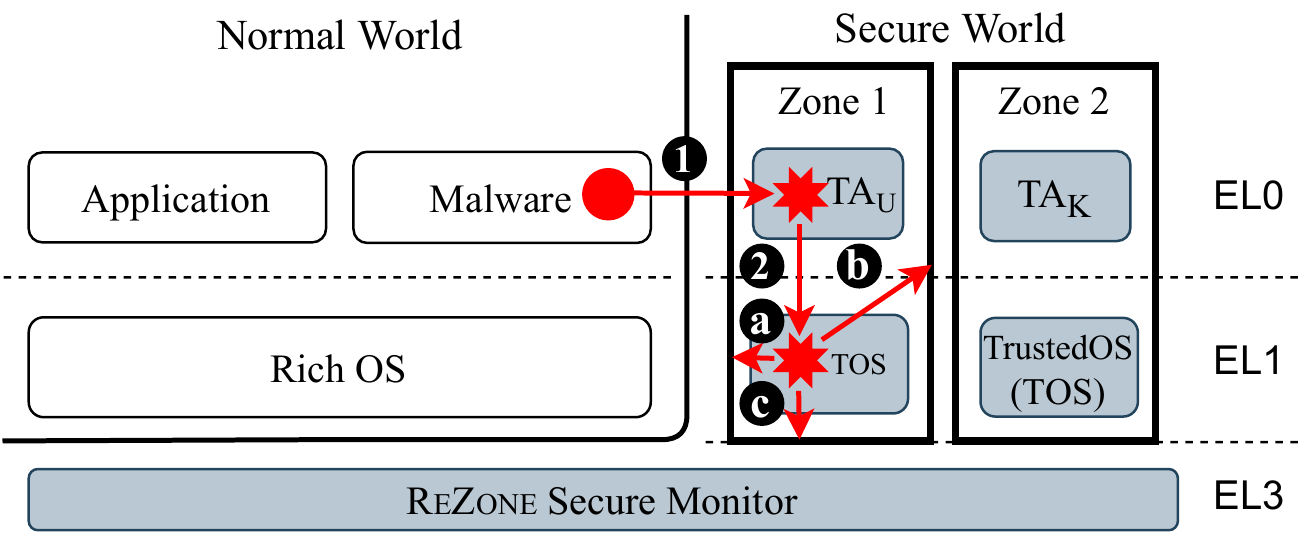}
    \vspace{-0.3cm}
    \caption{TEE privilege reduction with \project: an attacker cannot escalate privileges from the compromised trusted OS running in zone 1 to (a) the rich OS, (b) the kernel-level TA$_\textrm{K}$ in zone 2, or (c) the secure monitor.
    }
    \label{fig:intro_rezone}
    \vspace{-0.2cm}
\end{figure}

This paper presents \project, a new security architecture that can effectively counter ongoing privilege escalation attacks by reducing the privileges of a potentially compromised trusted OS. We build our solution specifically for restricting the privileges of S.EL1 on Armv8-A platforms featuring the protection modes displayed in Figure~\ref{fig:intro_trustzone}, and then discuss how our techniques can be employed for reducing the excess of S.EL2 privileges in Armv8.4-A and ensuing architectures. With \project, a typical monolithic TEE can be restructured and partitioned into one or multiple \textit{zones}, which consist of sandboxed domains inside the secure world. Zones have access only to private memory regions, and provide an execution environment for untrusted S.EL$_\textrm{1/0}$ code, i.e., trusted OS and TAs. \project restricts the memory access privileges of the code running inside a zone, preventing it from arbitrarily accessing memory allocated for: (a) the normal world REE, (b) other zones, and (c) the secure monitor. Figure~\ref{fig:intro_rezone} depicts a setup featuring two zones -- 1 and 2 -- each of them runs a trusted OS instance as in a library OS hosting local user- or kernel-level TAs. An attacker hijacking the trusted OS in zone~1 (as explained in Figure~\ref{fig:intro_trustzone}), will not be able to further escalate its privileges beyond its sandbox. Additional zones can be created, potentially hosting different trusted OS software.

A central novelty of \project's design is that we leverage TrustZone-agnostic hardware primitives available in COTS SoCs to restrict the privileges of S.EL1 (trusted OS) code. Existing systems such as Sanctuary~\cite{Brasser2019} have already leveraged the TrustZone Address Space Controller (TZASC) for restricting access permissions to various physical memory regions in the normal world. However, S.EL1 code can still revert these permissions, rendering the TZASC insufficient for zone isolation. Instead, we use a platform resource controller such as the Resource Domain Controller (RDC) existing in i.MX8MQ and i.MX8QM SoCs, which allows specific bus masters to control memory access permissions. In \project, we harness this feature along with a low-end microcontroller also available in the SoC to control these permissions, effectively creating private memory regions for each zone. RDC and microcontroller can then be used as building blocks for restricting the memory access privileges of a (potentially compromised) trusted OS, and enforcing strong zone isolation.

Based on this insight, we have built \project. Our system includes multiple non-trivial optimizations and fine-grained mechanisms to guarantee system-wide security (e.g., avoiding cache-based attacks based on cached content available for EL3 and S.EL1 code). To make our design generally applicable to various hardware platforms, we specify the general properties of SoCs that satisfy \project's requirements. We have fully implemented a \project prototype for the i.MX 8MQuad EVK, and integrated it with Android OS as REE and with OP-TEE as TEE. Our code is open sourced~\cite{rezonesite}.

We evaluated \project using microbenchmarks and real-world applications. Our evaluation shows that despite the existence of non-negligible performance overhead in microbenchmark programs, \project does not sensibly degrade the performance of applications such as crypto wallets or DRM media decoders. The DRM application, which requires frequent transitions between the normal world and the zone, preserves its ability to replay video at high frame rates. We have surveyed 80 CVE vulnerability reports and estimate that \project can thwart 86.84\% of potential privilege escalation attacks arising from exploiting these bugs.

%=======================================================================
%> [BACKGROUND]
%=======================================================================	
\section{Background and Motivation}

\subsection{Standard TrustZone Mechanisms}

We focus on TrustZone for Armv8-A architectures not featuring S.EL2. CPU cores can operate in one of two states: \textit{secure} and \textit{non-secure}. At exception levels EL0 and EL1, a processor core can execute in either of these states. For instance, the processor is in non-secure exception level 1 (NS.EL1) when running REE kernel code, and in secure EL1 (S.EL1) if executing the trusted OS. EL3 is always in secure state. To change security states, the execution must pass through EL3.

TrustZone extends the memory system by means of an additional bit called \textit{NS bit} that accompanies the address of memory and peripherals (see Figure~\ref{fig:RZ_HW}). This bit creates independent physical address spaces for normal world and secure world. Software running in the normal world can only make \textit{non-secure accesses} to physical memory because the core always tags the bus NS bit to 1 in any memory transaction generated by the normal world. Software running in the secure world usually makes only \textit{secure accesses} (i.e., NS=0), but can also make non-secure accesses (NS=1) for specific memory mappings using the NS flags in its page table entries. The processor maintains separate virtual address spaces for the secure and non-secure states. Trusted OS (or the secure monitor) can map virtual addresses to non-secure physical addresses by setting to 1 the NS bit in its page table entries.

TrustZone-aware memory controllers are usually configured to assign non-secure/secure physical address spaces to different physical memory regions. In the Arm architecture, the TZASC enforces physical memory protection by allowing the partitioning of external memory on secure and non-secure regions. (The TZPC performs a similar function for peripherals.) To date, Arm has released two TZASC versions: Arm CoreLink TZC-380 and TZC-400. They share similar high-level features, i.e., the ability to configure access permissions for memory regions (contiguous address space areas).

\subsection{The Excess of Privileges of the Trusted OS}
\label{sec:challenge}

The TrustZone mechanisms described above were designed under the assumption that the S.EL1 (trusted OS) and EL3 (secure monitor) are trusted. However, if the trusted OS is hijacked, the attacks described in Figure~\ref{fig:intro_trustzone} can be trivially mounted. For illustration purposes, suppose that the physical memory addresses PA$_\textrm{OS}$, PA$_\textrm{TK}$, and PA$_\textrm{SM}$ are allocated to the rich OS, TA$_\textrm{K}$, and secure monitor, respectively. The attacker could mount said attacks by creating virtual address mappings in the S.EL1 page tables as indicated below, and use the virtual addresses VA$_i$ to access (read/write) the memory pages of the rich OS, TA$_\textrm{K}$, or secure monitor. Notation VA$\rightarrow\langle$PA, NS$\rangle$ denotes a page table entry translating virtual address VA to the physical address PA and associated NS bit value:

\noindent VA$_\textrm{OS}\rightarrow\langle$PA$_\textrm{OS}$, 1$\rangle$, VA$_\textrm{TK}\rightarrow\langle$PA$_\textrm{TK}$, 0$\rangle$, VA$_\textrm{SM}\rightarrow\langle$PA$_\textrm{SM}$, 0$\rangle$

Simply put, \emph{the secure monitor cannot prevent the trusted OS from mapping these memory regions arbitrarily into the S.EL1 address space}. Given that the trusted OS does not require this level of control to sustain its typical operations, this means that S.EL1 is endowed with excessive privileges that can be abused. If the trusted OS is compromised, the attacker can leverage these privileges to take over every part of the system, including the monitor, TZASC, and normal-world components. For this reason, we argue that reducing the memory access privileges of S.EL1 code can significantly lower the impact of vulnerability exploits targeting the trusted OS. In this paper, we concentrate primarily on devising a solution to this problem. Then, in $\S$\ref{sec:sel2} and $\S$\ref{sec:cca}, we explain that Arm's architecture releases starting from Armv8.4 did not entirely solve the problem of over-privileged protection modes in the secure world, and discuss how our techniques can potentially be applied to these more recent architectures.

\input{TableCOTS}

\subsection{Platform Model for Restricting S.EL1}
\label{sec:plat_req}

\begin{figure}
    \centering
    \includegraphics[width=0.99\linewidth]{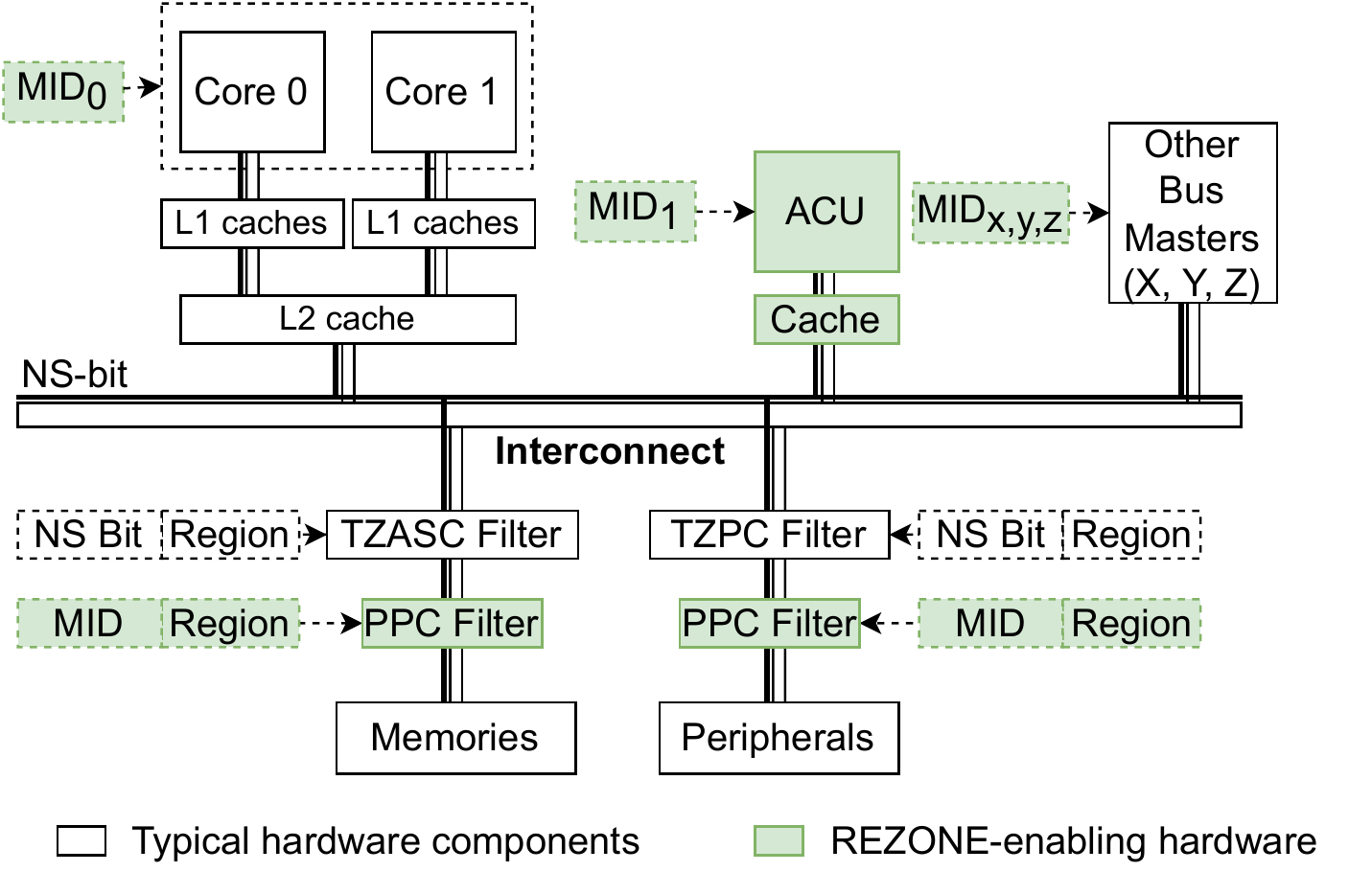}
    \vspace{-0.2cm}
    \caption{Hardware platform featuring TrustZone: \project-enabling hardware components are highlighted, i.e., ACU and PPC. PPC is configured by the bus master MID$_\textrm{1}$ (ACU). These configurations determine which access permissions are granted to a given MID for a specific physical memory region.}
    \label{fig:RZ_HW}
    \vspace{-0.1cm}
\end{figure}

Reducing privileges of S.EL1 code to prevent unauthorized memory accesses is not trivial using standard TrustZone mechanisms alone. Even though TZASC controllers such as TZC-380 and TZC-400 can restrict memory accesses, since S.EL1 can reconfigure the TZASC, the trusted OS can always override any permissions for blocking the trusted OS's access to certain memory regions. Also, it is not possible to distinguish accesses from different privilege levels, thus secure monitor (VA$_\textrm{SM}$) and TAs (VA$_\textrm{TK}$) memory cannot be restricted.

Given the limitations of vanilla TrustZone mechanisms, we propose to leverage complementary, TrustZone-agnostic hardware features available in COTS platforms for preventing illegal memory accesses potentially performed by untrusted S.EL1 code. These features consist of three hardware primitives that we require to be offered by the underlying platform.

\mypara{1. Platform Partition Controller (PPC).} Consists of a memory and peripheral protection controller that can restrict physical accesses to a given resource, i.e., memory or peripheral (see Figure~\ref{fig:RZ_HW}). On the i.MX8MQ SoC, the PPC is implemented by the RDC peripheral. We characterize the PPC by having four main properties. It must be capable of: (a) intercepting all physical requests regardless of both the NS bit signal injected into the system bus and the configurations of the TZASC, (b) setting fine-grained access control attributes (e.g., RO, RW) for configurable memory regions; (c) differentiating between multiple bus masters, and (d) protecting memory-mapped input/output (MMIO) regions, including the PPC's memory-mapped registers. Importantly, (c) and (d) help to ensure that the PPC's configurations cannot be arbitrarily changed by (untrusted) S.EL1 code executed by the processor. Intuitively, to provide this guarantee in the setup illustrated in Figure~\ref{fig:RZ_HW}, the PPC would not allow the bus master identified with MID$_\textrm{0}$ -- i.e., the processor cluster -- to arbitrarily change the PPC configurations. Instead, as explained next, it is the bus master identified with MID$_\textrm{1}$ that will have that privilege.

\mypara{2. Auxiliary Control Unit (ACU).} ACUs are small microcontrollers like Cortex-M4 added into the SoC for housekeeping purposes. For example, Figure~\ref{fig:RZ_HW}, the ACU is identified by MID$_\textrm{1}$. The ACU coordinates the configuration of memory access permissions in the PPC, ensuring that the main processor cores cannot tamper with the PPC permissions.

\mypara{3. Secure boot.} Secure boot validates the integrity and authenticity of the firmware, ensuring that the ACU has been securely bootstrapped and fully controls the PPC.

\subsection{Hardware Support on COTS Platforms}
\label{sec:PPC_ACU_COTS}

The primitives presented above are readily available on i.MX8MQ and i.MX8QM SoCs, the first of which we use for building \project. In i.MX8MQ, PPC and ACU are implemented, respectively, by RDC and Cortex-M4 hardware. To further assess if \project can be applied at a large scale, although secure boot is already commonly available, we need to determine if COTS platforms include hardware components that can play the role of PPC and ACU. To this end, we analyzed 17 popular SoCs from nine different vendors. We based our study on several sources. In some cases, we consulted publicly available SoC reference manuals. In others, we inspected the source code of the Linux kernel, bootloader, and secure monitor looking for PPC/ACU driver support. CVEs were also useful for identifying a PPC-like mechanism on Qualcomm SoCs. Our main findings are presented in Table~\ref{tab:rezone-cots}.

\mypara{PPC hardware candidates.} Apart from the RDC, we identified several other components that can be used as PPC: SMMU, XPU, and XMPU in conjunction with XPPU. The Arm SMMU is an IOMMU architecture that can translate an input address to an output address based on address mapping and memory attributes (i.e., translation tables). The main requirement for an IOMMU to be used as a PPC is the possibility of performing address translation functions for multiple masters, in particular the CPU. The Arm SMMU architecture, in its different versions, i.e., Arm SMMUv1, SMMUv2, SMMUv3, can fully provide such a feature. All of our studied platforms by Nvidia, Qualcomm, and Broadcom include a full-fledged SMMU. However, some custom SMMU implementations are designed to only interpose accesses of specific bus masters, e.g., media processing hardware. Such is the case of Mediatek 1200 and Kirin 9000. Samsung Exynos 900/2100 include dedicated SMMUs for specific peripherals. However, due to the lack of available information, we cannot ascertain that these SMMUs can also filter CPU accesses. The XPU~\cite{qualcommxpu} is a Qualcomm hardware component that entirely fills the PPC requirements. The same is true for Xilinx's XMPU and XPPU, which restrict accesses to memory and peripherals respectively~\cite{xilinxisol}. In general, we observe that recent platforms are more likely to feature PPC-like mechanisms\footnote{At a first glance, it may seem that TZASC fits all the PPC requirements enumerated in $\S$\ref{sec:plat_req}. However, a TZASC only protects memory regions. Therefore, it misses the PPC requirement (d). Additionally, only TZASC-400 can differentiate between bus masters, a highly specific hardware component not widely deployed. In fact, looking at the latest TF-A release~\cite{tf-a} only two platforms support TZASC-400: NXP's lx2160a and ls1028a. To control access to peripherals, the TZPC is also needed. However, available documentation~\cite{tzpc} shows that the TZPC cannot distinguish between multiple bus masters or prevent accesses to itself, missing PPC requirements (c) and (d).}.

\mypara{ACU hardware candidates.}
On the analyzed SoCs, we identified various ACU candidates. Qualcomm, Samsung, and NXP's i.MX8QM include security co-processors that can be used for this purpose: Secure Processing Unit (SPU)~\cite{SPU2019}, integrated/embedded Secure Element (iSE and eSE), and System Controller Unit (SCU). All other surveyed SoCs incorporate programmable microcontrollers usually reserved for platform management purposes, e.g., power management. With one exception, these microcontrollers can be used as a full-fledged ACU: Mediatek's SoC has a System Control Processor (SCP), but the SCP interface is based on MMIO registers and not a mailbox interface as the other analyzed ACUs. Therefore, this SoC is likely too restrictive for \project. In summary, our findings are consistent with Arm's reports~\cite{armscp}, which point to a clear trend in COTS SoCs to include companion processing units for power management and security operations.

\mypara{Platform support for \project.} In our study, we found that 13 SoCs have suitable PPC and ACU hardware, which make the deployment of \project possible on devices featuring these platforms. In contrast, the SoCs from Mediatek and HiSilicon lack at least one of the required PPC or ACU mechanisms to accommodate \project. In the case of Samsung's SoCs, we miss important information to issue a final judgment. In general, our study shows that our defense mechanisms can further be ported to platforms beyond the i.MX8 SoC family, which we use for our implementation. All NXP's i.MX8, Xilinx's Zynq, and Qualcomm's Snapdragon SoCs deployments span a wide range of edge computing settings (e.g. automotive)~\cite{qc_report}. It is interesting to note that most of the analyzed Qualcomm platforms are compatible with \project, which means that our system can potentially be deployed on many commodity mobile devices.

%=======================================================================
%> [Goals and Threat Model]
%=======================================================================	
\section{Design Goals and Threat Model}
\label{sec:threat_model}

We aim to design a security architecture that leverages the primitives presented above to create secure world sandboxes. We refer to these sandboxes as \textit{secure world zones}, or simply \textit{zones}. A zone is meant to host a partitioned TEE stack running at S.EL$_\textrm{1/0}$, i.e., trusted OS and TA software (see Figure~\ref{fig:intro_rezone}). Since multiple zones $z_i$ can co-exist within the secure world, we use the superscript notation S.EL$^i_\textrm{1/1}$ to identify zone $i$. More concretely, we strive to attain four subgoals:

\mypara{1. Reduce the privileges of the trusted OS.} As depicted in Figure~\ref{fig:intro_rezone}, zones must thwart the escalation of privileges of an attacker from the trusted OS to other memory regions of the system. To this end, a zone $z_i$ is expected to enforce three security properties: (P1) protection of the normal world, i.e., software running at NS.EL$_\textrm{1/0}$, (P2) protection of the secure-monitor, i.e., EL3 code, and (P3) protection of co-located zones, i.e., software running at S.EL$^j_\textrm{1/0}$, where $i\ne{j}$.

\mypara{2. Depend on a small TCB.} To implement zones in the secure world, we will rely on the secure monitor as root of trust, and incorporate as little additional code as possible into the TCB, e.g., for managing PPC and ACU.

\mypara{3. Maintain TEE software portability.} Legacy TEE and TA software abound. Vendors and OEMs should easily be able to port preexisting TEE/TA stacks into zone-based environments, requiring minimal to no changes in the code.

\mypara{4. Offer a good performance/security trade-off.} The overheads incurred by our solution should not significantly slow down typical real-world TrustZone-based TEE workloads. In other words, we are willing to trade some loss of performance, as long as the performance degradation is not detrimental to the user experience, for improved security guarantees.

\mypara{Threat model.} 
As for the threat model, the attacker's main goal is to subvert the security properties of zones (i.e., P1-3 shown above in subgoal~1). We assume the adversary has already compromised the trusted OS of a given zone $z_i$, and aims to escape it (see Figure~\ref{fig:intro_rezone}). Running in S.EL1$^i$, the attacker may attempt to access (read/write) external memory addresses outside the $z_i$'s private memory ($\S$\ref{sec:challenge}). These access requests may target i) the physical address space of the victim at NS.EL$_\textrm{1/0}$, EL3, or S.EL$^j_\textrm{1/0}$, or ii) the PPC or memory regions used by the ACU. In the first case, the objective is to override the PPC's memory-mapped registers containing the permissions that forbid access to the victim's memory regions. In the second case, the idea is to subvert the ACU's behavior, e.g., causing the ACU to hand over the control of the PPC to the processor core executing the attacker's code. This would indirectly allow the attacker to disable the PPC's restrictions. The attacker may also leverage the effect of caches to read or write cached memory cells for which S.EL1$^i$ does not currently have permissions.
 
In this work, we do not consider software attacks exploiting vulnerabilities in the secure monitor or the ACU software. We assume that these components are correct and belong to the system's TCB. The platform's secure boot initializes the system to a known state. We do not consider physical attacks that tamper with hardware, e.g., fault injection. Our solution has the side-effect of providing cache side-channel protection between zones. However, microarchitectural side-channels are out of the scope of our work. Similar to a typical TrustZone deployment, we do not consider denial-of-service (DoS) attacks including those caused, for example, by a malicious trusted OS not relinquishing control to the normal world.

%=======================================================================
%> [DESIGN]
%=======================================================================	
\section{Design}
\label{sec:overview}

\begin{figure}[t]
    \centering
    \includegraphics[width=0.98\linewidth]{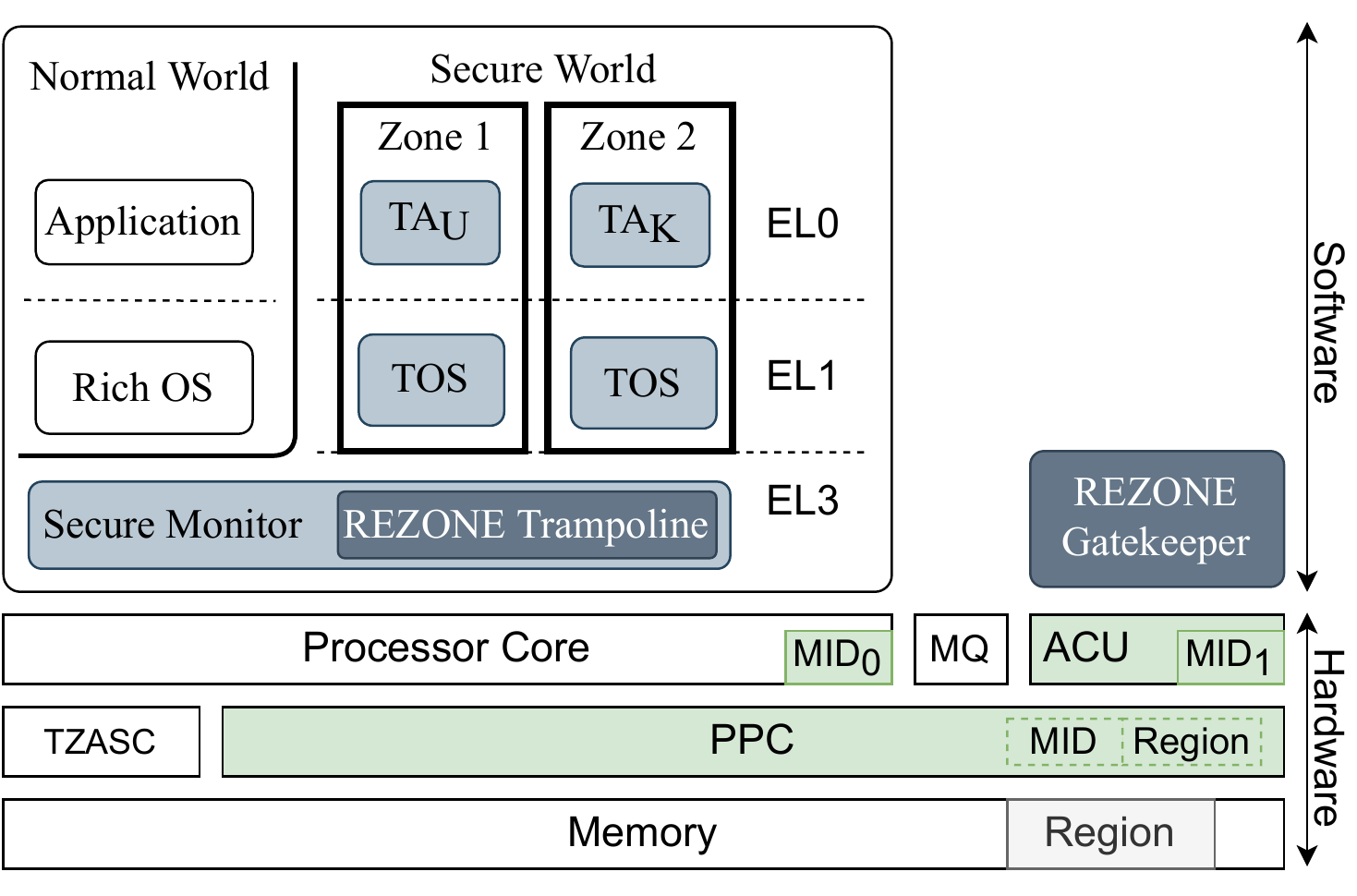}
    \vspace{-0.3cm}
    \caption{\project architecture: Its software components are colored in dark shade and consist of secure monitor (which includes the trampoline) and gatekeeper. Secure monitor runs on the main processor cores; gatekeeper runs on the ACU. \project software controls memory accesses via the PPC.}
    \label{fig:rezone_arch}
    \vspace{-0.3cm}
\end{figure}

Figure~\ref{fig:rezone_arch} represents the architecture of \project. Hardware-wise, our system relies on a typical TrustZone-enabled platform. For controlling memory access permissions, in addition to a TZASC controller, \project relies on a PPC hardware component. The PPC is dynamically configured to block secure world accesses from the processor based on the processor's bus master ID (MID$_\texttt{0}$). The PPC can be reconfigured only by a bus master, predefined at bootstrapping time. In \project, this bus master is the ACU (MID$_\texttt{1}$). ACU and processor can communicate with each other efficiently using a message queue (MQ) implemented by a hardware peripheral.

Software-wise, \project comprises the \textit{secure monitor} and the \textit{gatekeeper}. The former consists of standard secure monitor software (e.g., implemented by Arm Trusted Firmware) augmented with a \project-specific sub-component named \textit{trampoline}. The secure monitor (and trampoline) run on the main processor core and the gatekeeper on the ACU; the PPC protects their private memory regions, which store security-sensitive context information. Taken together, trampoline and gatekeeper manage the execution of zones in the system. They ensure that each zone can access only a private physical memory address space assigned to the zone, and take care of all context-switching tasks involving zone entering and exiting operations. These operations occur when an REE application makes a call to a zone's guest TA (\textit{zone entry}), and the TA returns the results of the call (\textit{zone exit}). REE and zone can share data through a shared memory region. \project's software components are shipped with the platform firmware. When the system bootstraps, the firmware configures the memory layout and statically creates one or multiple zones indicating the composition of their respective software stacks, i.e., trusted OS and TAs. Next, we explain our design decisions and how the system works. We refer the reader to Figure~\ref{fig:HZ-exec-model}.

\subsection{Memory Partitioning and Permissions}
\label{sec:memory_restrictions}

\begin{figure*}[!ht]
    \centering
        \includegraphics[width=0.99\linewidth]{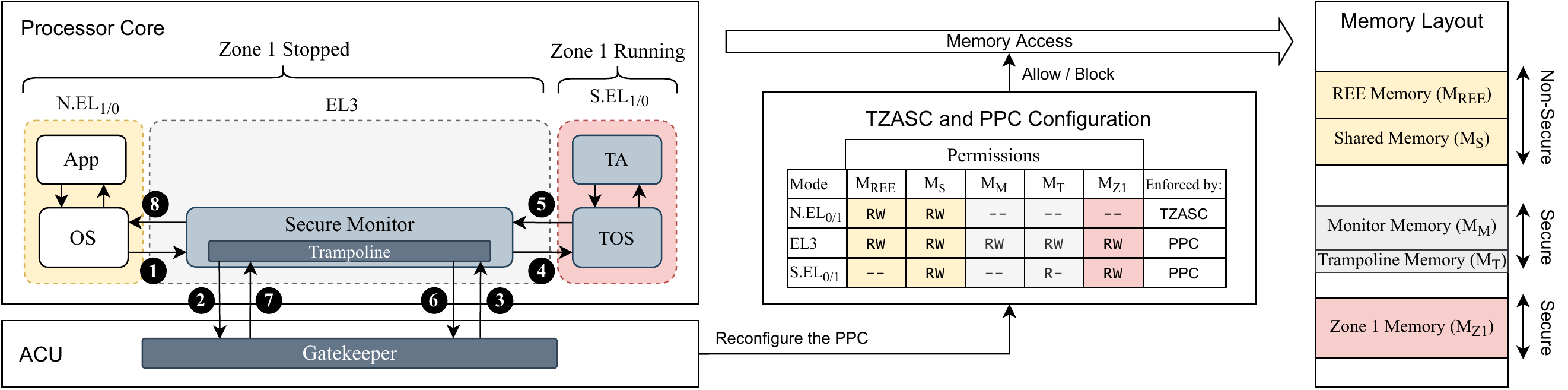}
    \vspace{-0.3cm}
    \caption{\project execution model, memory access permissions, and memory layout: \project ensures that the code running inside zone 1 -- i.e., at S.EL$_\textrm{1/0}$ -- can only access zone 1's private memory region (M$_\textrm{Z1}$). This protection is achieved by reconfiguring the PPC to restrict the processor's access permissions when entering zone 1 (steps 3\&4), and reverting these configurations when exiting zone 1 (steps 5\&6). %The PPC is used only to restrict the secure accesses.
    }
    \label{fig:HZ-exec-model}
    \vspace{-0.3cm}
\end{figure*}

The starting point in \project is to partition the physical memory layout into regions, allocating these regions to different protection domains of the system, and specifying regions' memory access permissions accordingly. In doing this, we are primarily concerned about restricting secure world accesses when the processor runs zone code, i.e., S.EL$_\textrm{1/0}$. The final memory layout and access permissions are depicted in Figure~\ref{fig:HZ-exec-model}. We progressively explain these regions and permissions, beginning with a vanilla TrustZone-enabled platform, and then extending it with \project's protections.

Just like in a typical TrustZone setup, in \project, the normal world includes regions for the REE OS (M$_\texttt{REE}$) and shared memory (M$_\texttt{SH}$). The secure world has independent regions for the TEE software. The TZASC is then configured to prevent the normal world software from accessing secure world regions. The first line in the permissions table (see Figure~\ref{fig:HZ-exec-model}) shows how the normal world has only read/write access permissions to normal world memory regions.

Then, we further separate the secure world memory into independent regions for the secure monitor and for each zone $i$, e.g., zone 1 (M$_\texttt{Z1}$). We need to ensure that the secure monitor software (EL3) has full access permissions, but the code running at S.EL$_\textrm{1/0}$ is confined to (i) the zone's private memory M$_\texttt{Z1}$ and (ii) the REE shared memory for TA calls. To this end, we leverage the PPC to restrict secure world accesses. However, using the PPC for this purpose is not trivial. Given that TrustZone does not mark memory accesses with the privilege level from which they originate (being tagged per world only) the PPC cannot distinguish secure accesses coming from EL3 or from S.EL$_\textrm{1/0}$. As a result, PPC's permissions need to be dynamically changed by the monitor when entering a zone (i.e., reprogramming the PPC to remove \texttt{RW} access permissions to M$_\texttt{REE}$ and to secure world memory as per line three in the permissions table) and when leaving a zone (i.e., reverting to the permissions in line two of the table).

However, a difficulty arises when exiting a zone. If the S.EL1 code running inside a zone is wholly prevented from accessing the secure monitor memory, this protection will still be in place by the time the zone's trusted OS traps into EL3; consequently, the secure monitor will be blocked by PCC resulting in memory access aborts and a system crash. To solve this problem, we create a dedicated secure memory region containing code and data that are necessary only to facilitate this transition securely. This code is part of the trampoline, which still belongs to the secure monitor because it runs in EL3. We then segregate a region of the trampoline's memory (M$_\texttt{T}$) from the rest of the secure monitor memory (M$_\texttt{M}$) because these regions need to be configured with different memory access permissions. M$_\texttt{M}$ is always denied access from S.EL$_\textrm{1/0}$. In contrast, M$_\texttt{T}$ is marked read-only by the PPC, therefore, allowing the processor to both execute trampoline code from M$_\texttt{T}$ and load context data also from M$_\texttt{T}$ while still running in S.EL1. The trampoline code will then engage the ACU to remove the PPC memory access restrictions to M$_\texttt{M}$, thus allowing the secure monitor to freely run at S.EL3 and restore the normal world execution. These reasons justify the PPC permissions in the last two rows of the table in Figure~\ref{fig:HZ-exec-model}.

\subsection{Securing the PPC Memory Permissions }
\label{sec:PPC_memory_restrictions}

To guarantee the enforcement of the memory access permissions as described above, the PPC must be (re)configured exclusively by a trusted component. In \project, this component is the gatekeeper/ACU ensemble. We configure the PPC so that only the ACU is authorized to perform this operation. Since the PPC can differentiate requests based on the bus master ID, we program the PPC to accept configuration requests only by a bus master identified with the ACU's MID. Requests originating from other bus masters, e.g., the processor, will be denied. The ACU runs the gatekeeper code and dynamically updates the PPC permissions when entering or leaving a zone by serving the specific requests sent by the trampoline, e.g., to enable M$_\textrm{M}$ \texttt{RW} permissions at zone exits.

\project also needs to defend against a potential attack. Processor and ACU communicate via a message queue unit (MQ in Figure~\ref{fig:rezone_arch}) which is agnostic to the current exception level of the processor. As a result, an attacker running at S.EL1 in a zone could craft a request to disable the memory access restrictions enforced by the PPC. To thwart this threat, the gatekeeper must foretell that this request is malicious and therefore refuse it. To this end, trampoline in EL3 and gatekeeper share a secret token that is initialized at boot time and copied to their private address space. This token is then used by the trampoline for authenticating PPC reconfiguration requests sent to the gatekeeper. Since the secret token is not accessible at S.EL1, an attacker can no longer spoof legitimate requests, making the system robust against these attacks.

\subsection{Preventing Cross-zone Interference}
\label{sec:cross_interference}

The mechanisms discussed so far secure all memory transactions observed outside the processor cluster according to the permissions matrix in Figure~\ref{fig:HZ-exec-model}. These mechanisms require additional measures to safeguard against potential attacks within the processor itself. We address two concrete concerns.

First, the PPC acts only at the system bus level. Therefore, its memory access protections do not extend to the internal caches of the processor cluster. For this reason, if the caches contain data that a given zone is not authorized to read or write, and the core enters that particular zone, the content will be accessible to that zone, which constitutes a security violation. To prevent these problems, the cache hierarchy of the core executing the zone is flushed by the trampoline.

Second, the PPC cannot differentiate between individual cores within a cluster, because the PPC observes requests as being from a single bus master: the cluster's shared cache. As a result, \project's memory restrictions enforced by the PPC are collectively applied across all cores of the same cluster. However, blocking access to the normal world memory when a core enters a zone will also prevent other co-located cluster cores from accessing normal world memory, which may lead to crashes if they are currently executing REE code. To prevent this problem, upon zone entry, \project must halt other co-located cluster cores running in the normal world until the core exits the zone. Despite the non-negligible performance overheads potentially introduced by these measures, our experiments show that these overheads can be tolerated due to the infrequent TEE invocation patterns and short TA execution times inside the secure world (see $\S$\ref{sec:eval}), essentially trading some performance degradation for increased TEE security.

\subsection{Zone Entry and Exit Workflows}
\label{sec:key_mechanism}

The protection mechanisms described above act together whenever an REE application invokes a TA hosted in a zone.

A zone entry begins when the normal-world OS issues a request through the invocation of the \texttt{smc} instruction (1), causing the execution to be handed over to the secure monitor. If the request needs to be handled by the trusted OS, \project will first differentiate to which zone the requests to that particular TA have been statically assigned. Then, it forwards the request to the respective zone, i.e., zone 1 in Figure~\ref{fig:HZ-exec-model}. For this, the secure monitor starts by notifying all other cores in that cluster. The other cores will then enter an idle state. The trampoline first flushes the core's and shared cache content and then continues by issuing a request (2) to the gatekeeper to configure the PPC to block access of the application core. To authenticate the request as coming from the secure monitor processor mode, the gatekeeper expects to receive the secret token, which was handed to the monitor during boot. After a successful reply from the gatekeeper (3), the trampoline will then (4) jump to the trusted OS.

A zone exit occurs once the trusted OS returns execution to the secure monitor (5). At this point, the trampoline issues a new request to the gatekeeper (6) to unblock secure memory accesses from the core. The gatekeeper ensures this by reconfiguring the PPC to disable the zone's protection policies, i.e., granting \texttt{RW} for all memory regions. The secure monitor will then (7) resume execution, notify other cluster cores to, finally, return to the normal world context (8) executing under traditional TrustZone restrictions enforced by the TZASC.

\newcounter{problem}
\newenvironment{problem}[1]{\refstepcounter{problem}\par\medskip
   \textbf{Problem~\theproblem. #1: } \rmfamily}{\medskip}
   
%=======================================================================
%> [IMPLEMENTATION]
%=======================================================================	
\section{Implementation}
\label{sec:impl}

The implementation of \project is tightly dependent on the targeted hardware. We used the i.MX 8MQuad Evaluation Kit (EVK), which features an i.MX8MQ SoC by NXP. This SoC is powered by (i) one cluster of four Cortex-A53 application processors (@1.5 GHz) and (ii) one microcontroller, a Cortex-M4 (@266 MHz). The Cortex-A53 cluster has 32KiB level-1 (L1) instruction and data caches per core and a shared 2MiB level-2 (L2) cache. The Cortex-M4 has a 16KiB instruction cache and a 16KiB unified cache.

We use the Cortex-M4 to act as \project's ACU, and an SoC-embedded controller named RDC to play the role of PPC. This controller enables fine-grained control of access permissions through the creation of security domains. There are at most four domains, each identified by an ID (DID$_i$). A domain represents a collection of platform resources (e.g., memory regions, peripherals), and it is tagged with a set of per-resource access permissions (e.g., read/write, read-only). Bus masters, such as the processor cluster, microcontroller, DMA, etc., can be associated with one or multiple domains. A bus master in DID$_i$ can only access a resource if the resource is also associated with the bus master's DID$_i$, and its access privileges will be determined by the permissions of DID$_i$.

Based on the security domain scheme provided by this PPC, we use the following insight to implement the memory partitioning and access permissions described in $\S$\ref{sec:memory_restrictions}. Since the PPC is itself a platform resource, we need to reserve a domain (e.g., DID$_1$) with full read/write permissions to the PPC's registers, and associate the ACU as the single bus master to DID$_1$. Doing that, we guarantee no other bus master in the system will be able to tamper with PPC's security permissions. If DID$_1$ can be specified on the system prior to any tampering from an adversary, we ensure that the ACU remains in full and exclusive control of the PPC. To achieve this, we can rely on the secure boot to run a customized initialization routine (i.e., gatekeeper code) that will fully initialize the ACU before starting the zones' trusted OSes.

Our implementation configures two security domains when booting the system: DID$_\textrm{0}$, and DID$_\textrm{1}$. DID$_\textrm{0}$ contains all memory (and other) resources exposed to the cluster. DID$_\textrm{1}$ mainly contains the PPC and the gatekeeper's private memory region. The processor is associated with DID$_\textrm{0}$, and the ACU to DID$_\textrm{1}$. As a result, only the ACU can reconfigure the PPC as per the access permission matrix depicted in Figure~\ref{fig:HZ-exec-model} (last two rows). In the i.MX 8M family, Cortex-A clusters are assigned a unique master ID. This means that DID$_\textrm{0}$'s permissions apply indistinguishably to every Cortex-A53 core of the cluster.

We need to establish a secure channel between the cluster and ACU. Upon zone entry/exit events, they exchange message requests (reconfiguration) via the platform's MQ (see Figure~\ref{fig:rezone_arch}). In the i.MX8MQ SoC, the MQ is composed of MU\_A and MU\_B. The PPC is configured to assign MU\_A exclusively to DID$_\textrm{0}$, and MU\_B to DID$_\textrm{1}$, ensuring that only the processor and ACU can communicate with each other through the local MU interfaces. To authenticate the requests sent by the processor, trampoline and gatekeeper share a secret token. To preserve the secrecy of the token, the trampoline stores it in the \texttt{TPIDR\_EL3} 64-bit register, which is exclusively available to EL3 code (not even S.EL1 code can read this register) leaving the attacker a 1 in $2^{64}$ chance of guessing its value correctly. The token is randomly generated at boot time.

Given that the ACU runs at a clock speed 5.6$\times$ slower than the cluster, we implemented an optimization to offload workload from the gatekeeper (ACU) to the trampoline (cluster). We observed that the reconfiguration of the PPC's permissions DDI$_\textrm{0}$ resources become slower as the number of resources grows. To shift most work to the trampoline, the gatekeeper temporarily associates the PPC to DDI$_\textrm{0}$, allowing the trampoline to update PPC's permissions on the gatekeeper's behalf. Once finished, gatekeeper removes PPC from DDI$_\textrm{0}$, reacquiring its former condition of sole guardian of the PPC.

\subsection{Cross-core Synchronization}
\label{sec:synch}

While one core is in a zone (secure world), the PPC blocks access from the whole cluster to several memory regions (e.g., normal world memory). Thus, specific actions need to be taken into account to prevent other cores from hanging and crashing. Below are the instances where such scenarios might occur (for the sake of simplicity we consider only two cores):

\begin{enumerate}[wide, labelindent=0pt]
  \setlength\itemsep{0em}
    \item\label{sync1} Two cores $c_1$ and $c_2$ execute secure monitor code at EL3, and $c_1$ issues a zone entry, requesting a shift to S.EL1 while $c_2$ still runs at EL3. This can be a problem because if $c_1$'s request is attended without proper synchronization, $c_2$ will not be able to keep accessing monitor memory.
    
    \item\label{sync2} One core ($c_1$) is executing the trusted OS at S.EL1 while another ($c_2$) is sleeping. Meanwhile, the sleeping core $c_2$ awakens into monitor code (e.g. due to an interrupt). Again this is a problem because $c_2$ will not be able to access monitor memory while $c_1$ executes the trusted OS.
    
    \item\label{sync3} Two cores $c_1$ and $c_2$ are executing normal world code, and $c_2$ requires a service from a TA hosted inside a zone. This is a problem because, when $c_2$ enters the zone, $c_1$ will be denied access to normal world memory.
\end{enumerate}

We solve this problem by synchronizing the cores using spin locks and IPI interrupts. To address scenario \ref{sync1}, the monitor tracks which cores are currently executing the secure monitor. If a core intends to enter a zone, they all wait on a barrier when leaving EL3. At that point, $c_2$ sits on a spin lock while $c_1$ is allowed to jump into the trusted OS. Once $c_1$ concludes the zone call and returns to EL3, $c_2$ can continue. We solve scenario \ref{sync2} by waking the core $c_2$ into a trampoline's read-only memory region (not-writable by the running trusted OS) and waiting for the trusted OS in $c_1$ to cease execution by using a spin lock (\texttt{rz\_lock}). 
To solve scenario \ref{sync3}, we use an IPI interrupt and a spin lock (\texttt{rz\_lock}). First, when $c_1$ issues a request from normal world to enter a zone, it traps into the monitor (EL3) where it fires an IPI interrupt and grabs the \texttt{rz\_lock}. The goal of this interrupt is to notify other cores (i.e., $c_2$) that $c_1$ wishes to enter a zone. Core $c_2$ is then interrupted and jumps into trampoline code, an exception handler running in EL3, where it unlocks $c_1$ letting it continue to enter the zone; $c_2$ now waits on the spin lock until $c_1$ exits the zone, allowing $c_2$ to resume its execution in the normal world.

\subsection{Microarchitectural Maintenance}
\label{sec:micromaintenance}

\begin{figure}
    \centering
    \includegraphics[width=0.99\linewidth]{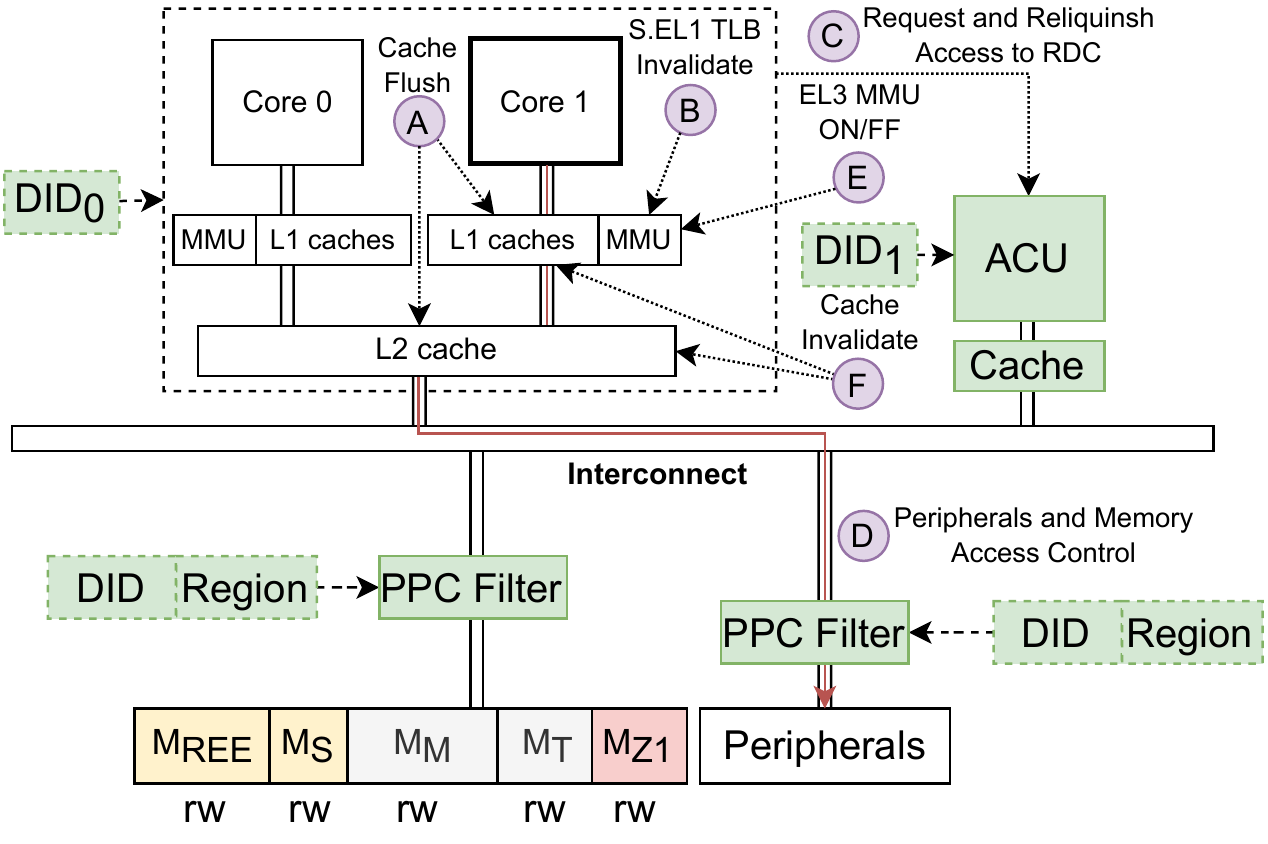}
    \vspace{-0.2cm}
    \caption{Context-switch between EL3-S.EL1 requires microarchitectural maintenance operations (A, B, E). Entering a zone (EL3$\rightarrow$S.EL1) requires A, B, C, D and E. Exiting a zone (S.EL1$\rightarrow$EL3) requires C, D, E and F.}
    \label{fig:scr_ctx_hw}
	\vspace{-0.3cm}
\end{figure}

Triggered by a zone entry/exit event, context switching between EL3-S.EL1 requires the trampoline to coordinate not only cross-core synchronization, but also perform important cleanup tasks (see Figure~\ref{fig:scr_ctx_hw}) and PPC reconfiguration operations. We now focus on microarchitectural maintenance that needs to be carried out, 
and then in $\S\ref{sec:dynamic-conf}$ we describe how the trampoline dynamically reconfigures the PPC.

When entering a zone (EL3$\rightarrow$S.EL1), the trampoline running on the core needs to ensure that the trusted OS (after the switch is finished) cannot access resources outside of the zone. After taking all measures to prevent interference from other cores ($\S$\ref{sec:synch}), the trampoline needs to invalidate the L1 instruction and data caches and the shared L2 cache (A). Unless this operation is performed, data or instructions left in the cached will be accessible to the trusted OS, thus violating zone isolation. Also, right before jumping into the trusted OS, the trampoline disables SMP coherency. Recall that at this point, all other cores in the cluster are suspended. Each core has data and instruction in its own L1 caches. With SMP coherency enabled, a request to access cached data would cause the coherency protocol to fetch the data from another core's L1, resulting in a cache hit; thus, data could be accessed by the present core, violating the zone's isolation properties.

In addition to cache maintenance, when switching between the trusted OSes of different zones, the trampoline needs to perform TLB maintenance (i.e., TLB invalidation) for the current core (B). The main reason is not due to concerns about violation of zone isolation, as this is already enforced by the PPC by setting each zone's memory region to be private. However, given that (i) the TLB for S.EL1 is shared by the trusted OSes of active zones, and (ii) the TLB has no way to tell which of the cached page table entries belong to each zone  (unless a TLB invalidation is performed), a TLB lookup by the trusted OS could translate into a physical address that belongs to another zone, thus causing unexpected faults. As an optimization, we perform TLB maintenance only when switching between different trusted OSes, and it does not affect TLB entries of the REE OS. With all microarchitectural maintenance done, the trampoline will restore the general-purpose register state and exit to the trusted OS.

\subsection{Dynamic PPC Reconfiguration}
\label{sec:dynamic-conf}

At the center of the context switch operation between EL3-S.EL1 is the dynamic configuration of PPC's memory access permissions. PPC reconfiguration guarantees that the trusted OS cannot access resources unless they are attributed to its zone. Specifically, the trampoline reconfigures the PPC to prevent S.EL1 code running inside a zone from accessing resources belonging to normal world, monitor, and other zones.

The reconfiguration of PPC's permissions is executed by the trampoline in EL3 after performing micro-architectural maintenance, as doing so before would prevent normal world and monitor memory from being written back to main memory. To perform the reconfiguration, the trampoline must first obtain access to the PPC. This step is achieved by sending a request to the ACU with the secret token to authenticate the trampoline (step C in Figure~\ref{fig:scr_ctx_hw}). The ACU will then allow the application cores to access the PPC. At this point, only the trampoline can access the PPC. The trampoline can then perform the necessary PPC configurations (step D) as per the permissions matrix presented in Figure~\ref{fig:HZ-exec-model}. After the PPC reconfiguration is concluded, the trampoline sends another request to the ACU. This time the ACU configures the PPC to prevent any accesses to the PPC by any bus master other than the ACU itself. Once the ACU replies to the trampoline, control can be safely handled over to the trusted OS.

\subsection{Handling Exceptions at S.EL1 Exits}\label{sec:safe_exit}

In contrast to zone entries, exiting a zone from S.EL1 to EL3 precludes steps A and B (see Figure~\ref{fig:scr_ctx_hw}). Flushing caches (A) is not necessary, since we selectively invalidate monitor memory in step F (explained below), thus preventing malicious cache manipulation to gain code execution privileges. TLB maintenance (B) is not required because S.EL1 TLB entries only affect S.EL1.  However, we need special precautions in dealing with exceptions from S.EL1. Two problems may arise:

\begin{enumerate}[label=\textbf{\arabic*.}, wide, labelindent=0pt]
  \setlength\itemsep{0em}
    \vspace{-0.2cm}
    \item\label{prob1} \textbf{TLB-miss:} Upon entering the vector table to handle an S.EL1 exception in EL3, if the TLB misses a page table entry that maps to the memory region of the exception handler, then the MMU must perform a page table walk. As the caches have been flushed before entering S.EL1, this operation will trigger an access to main memory to load the page table entry. However, since the EL3 page tables are stored in the secure monitor's memory region, whose access by EL3 code is still blocked by the PPC at this stage (recall only the trampoline memory is marked read-only), the code for handling the exception cannot be loaded and the system will hang.
    
    \item\label{prob2} \textbf{EL3 code injection from S.EL1:} From 1) it follows that the code on the EL3 vector table must be marked as read-only by the PPC, allowing also the trusted OS to read the handler code from memory. However, since the vector table code will be cached in the L1/L2 caches, write operations are still permitted to cache lines tagged as secure, as long as they do not reach the main memory (note the PPC acts only at the bus level). An attacker with S.EL1 privileges can use this feature to modify the exception handler code to its advantage, e.g., by modifying the SMC handler, an SMC call can trigger the secure monitor to load malware from the cache.
\end{enumerate}

To solve both problems, we disable the MMU for EL3 code (E in Figure~\ref{fig:scr_ctx_hw}). We do this in the last steps of exiting from EL3 to S.EL1 when entering a zone. Note that this measure only applies to the MMU for EL3 code: we do not disable the MMU for S.EL1 code. This solves problem 1, as, from that point on, memory accesses at EL3 will not be subject to MMU translation; thus, no page table accesses are needed, and the exception handler can then be executed. Disabling the MMU will also disable L1 and L2 caches for EL3 address spaces, and this will partially solve problem 2 as the exception handling code will always be fetched from main memory. This memory is protected read-only by the PPC, hence no modifications will be possible; however, this solution does not entirely solve the whole problem. Note that during the switch from EL3 to S.EL1, after flushing all caches, the trampoline still executes with caches enabled. This will inevitably leave some traces of trampoline data and instructions in the caches, which a malicious trusted OS can try to leverage for an attack by modifying them locally while running in S.EL1 as described above. To clean up potentially tampered cache content and avoid these attacks, we invalidate on-cache trampoline-related memory when the exception handler in EL3 is executed on an S.EL1 exit (F in Figure~\ref{fig:scr_ctx_hw}). After this cache management operation completes, the exception can be handled securely.

\subsection{Software Implementation}
\label{imp:sw}

The \project software is a firmware bundle based on TF-A (v2.0). TF-A implements basic bootstrapping and secure monitor functionality. We modified TF-A to incorporate the trampoline, encompassing 303 and 1220 SLOC of C and assembly code, respectively. We also used the TF-A to bootstrap the gatekeeper. The gatekeeper is implemented as a bare-metal application that runs on the Cortex-M4. It was written in 417 SLOC of C code and packaged with essential drivers to interface with the RDC. The system image was compiled with the GNU Arm toolchain for the A-profile Architecture (v9.2.1) and GNU Arm Embedded Toolchain (v9.3.1).

We also changed the TF-A's context management routines and data structures to support two zones, each running a trusted OS instance based on OP-TEE (v3.7.0). To run OP-TEE inside a zone, no intrusive modifications were necessary; we have disabled only the support for dynamic shared memory. We create the second OP-TEE instance in a different address space and updated the values of SMC IDs to be uniquely attributed to each OP-TEE instance. We created two full-blown stack builds featuring different REE: one running the Linux kernel (5.4.24) and the second the Android OS (AOSP 10.0). It was not necessary to modify REE code to support \project; however, since we added support to run multiple trusted OSes, we added drivers to interface with both of them. We also duplicated the OP-TEE Linux kernel driver and modified the SMC\_IDs used to make calls to the second OP-TEE instance. There are two tee-supplicant services, one for each OP-TEE.

%=======================================================================
%> [EVALUATION]
%=======================================================================
\section{Performance Evaluation}
\label{sec:eval}

We evaluated \project using the i.MX8 platform and software described in $\S$\ref{sec:impl}. On the i.MX8 platform, all cores, i.e., the Cortex-A53 APU cluster (4x) and the Cortex-M4, were running. Our performance evaluation covers three vectors:

\mypara{1. Microbenchmarks.} We evaluate the performance overheads at increasing levels of REE-TA interaction. Firstly, we measure the \textit{world switch time}, i.e. the time to transition from normal world to the first instruction of the trusted OS. Secondly, we measure the \textit{round-trip execution time of the GlobalPlatform TEE Client API} (v1.0), i.e., the standard API for an REE application to interact with the TEE. We tested 9 APIs. Lastly, we gauge the \textit{execution time of OP-TEE xtest}, a suite of regression tests provided by OP-TEE to verify the correct implementation of the trusted OS and related APIs. The suite has a total of 95 tests, grouped into nine groups. The suite also has seven benchmarks to assess the performance of cryptographic- and storage-related operations.

\mypara{2. Real-world applications.} We evaluated two TAs: a Bitcoin wallet and a DRM player service. The open-source Bitcoin wallet TA~\cite{bitcoinwallet} provides services, e.g., for creation and deletion of a master key, access the mnemonic for a master key, sign a transaction, and get the address of the wallet. The DRM player service is implemented with the open-source OP-TEE Clearkey plugin~\cite{opteeclearkey} and the media player ExoPlayer~\cite{exoplayer}. This service approximates to a security level 2 (L2) DRM compatible setup~\cite{widevinelevels}, where video content is split into subsamples and decrypted within the TEE, but processed in the normal world. The L2 DRM setup represents the worst-case scenario for performance since the process is split among the two worlds. We measured the (i) execution time for decrypting one subsample for three different resolutions and frame rates and (ii) the time elapsed between decoded content.  

\mypara{3. Impact on REE performance.} We used PCMark for Android benchmark (v3.0), which assesses the performance of smartphones by testing everyday activities. The suite consists of five real applications (e.g., web browsing, data and video editing) and it does not invoke secure world functionality. The benchmark provides an overall score (Work 3.0) translating the performance of the whole system into a quantifiable scoring system: a higher score indicating better performance.

\mypara{Methodology.} To measure world switch time, we used the AArch64 generic timer, a 64-bit cycle counter running at 8.33MHz. For the GlobalPlatform API and xtest suite, we used the Linux time API with the CLOCK\_REALTIME parameter enabled. For the Bitcoin wallet, we used the Linux time utility. For DRM service, we used (i) the Linux time API for the subsamples decryption and (ii) the Java System.nanoTime for the processing. Excepting world switch measurements, for all other experiments, we compare three system configurations: (i) standard OP-TEE deployment (No RZ); (ii) \project deployment (RZ); and (iii) \project deployment with no trusted OS preemption from the REE (RZ no IRQ). We collected 1000 samples and present the average of values within the 95$^{th}$ percentile, which removes the effect of occasional Linux interference. For microbenchmarks, we flush the caches at the end of each test. This reduces the hot-cache effect of the previous iteration and approximates a realistic TEE setup.

\begin{figure}[t]
    \centering
    \includegraphics[width=\linewidth]{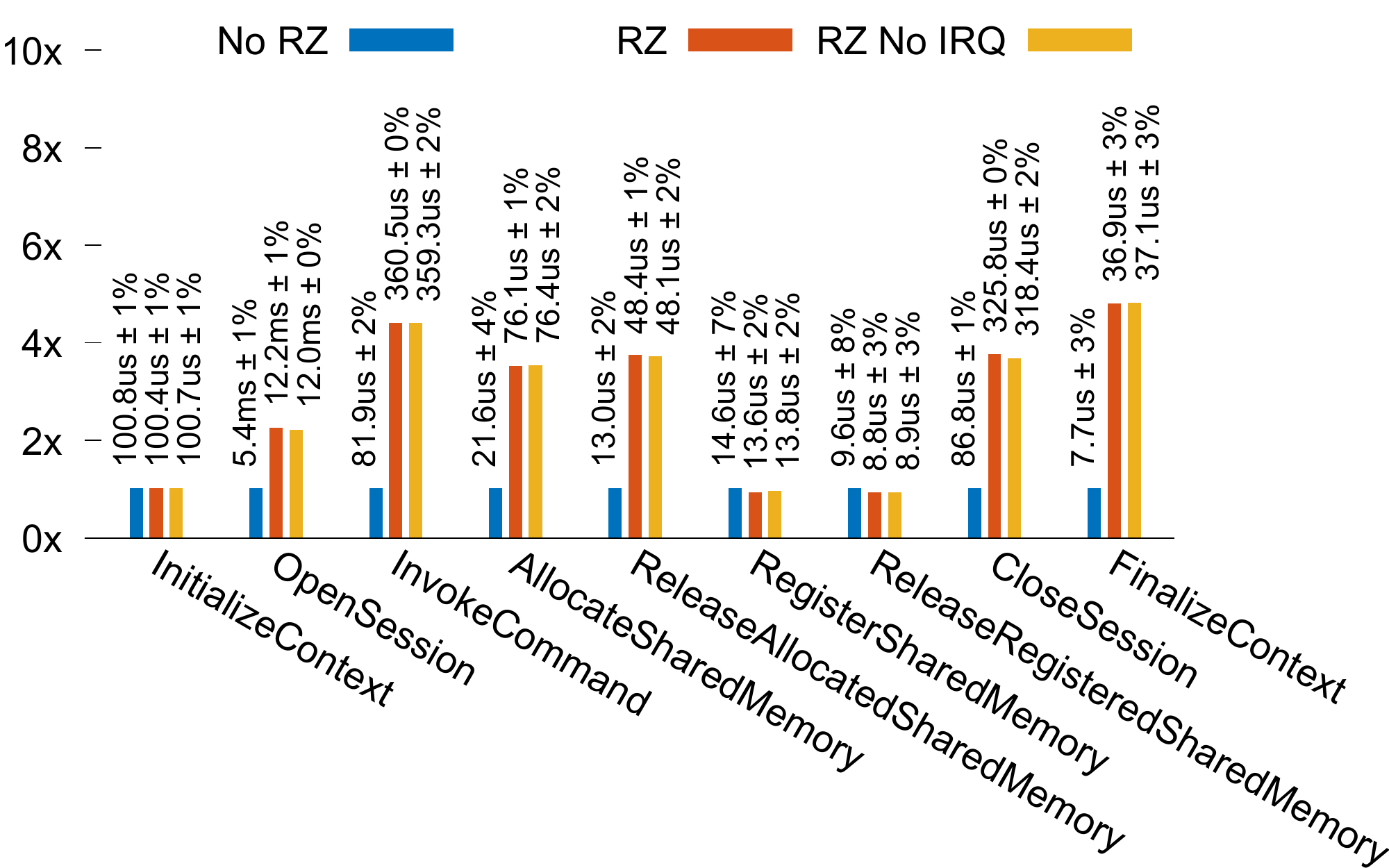}
    \vspace{-0.8cm}
    \caption{\project overhead, time, and normalized standard deviation (coefficient of variance) of GlobalPlatform API.}
    \vspace{-0.0em}
    \label{fig:api}
	\vspace{-0.3cm}
\end{figure}

\begin{figure*}[!t]
    \centering
    \includegraphics[width=0.9\linewidth]{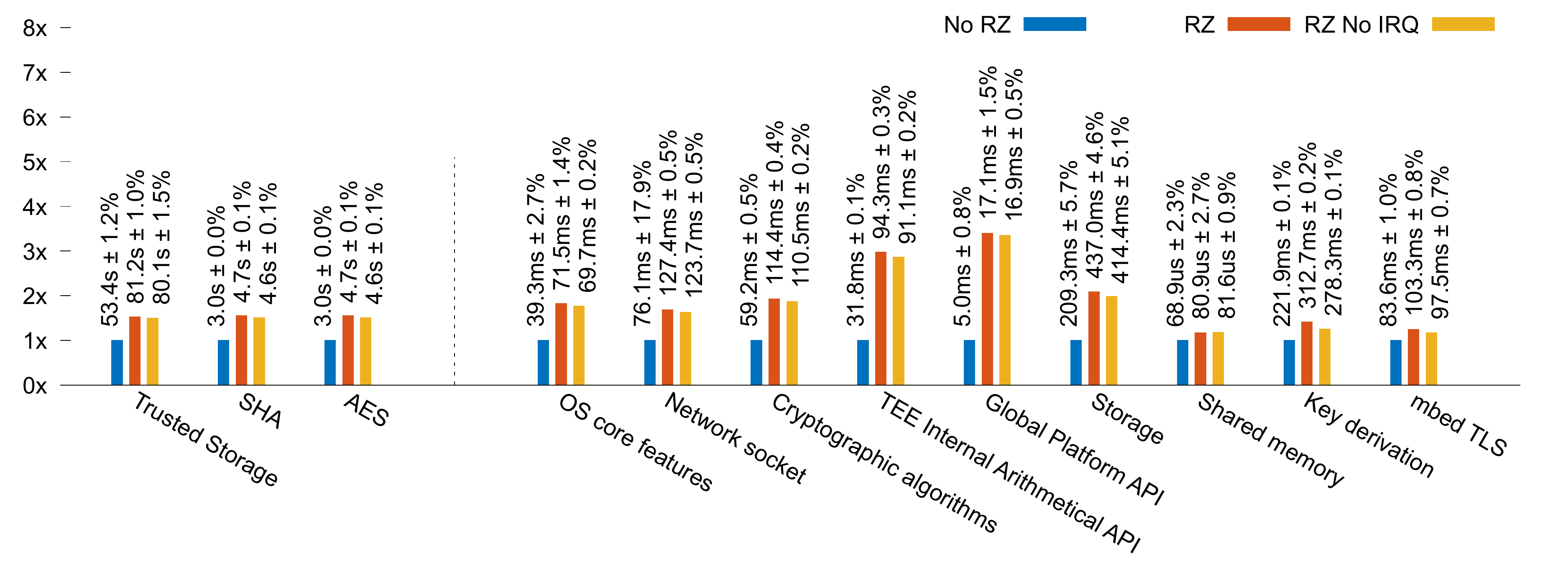}
    \vspace{-0.3cm}
    \caption{Geometric mean of \project overhead for xtest. Benchmarks (left) and unit tests (right). %Geometric mean is 1.74$\times$.
    }
    \label{fig:xtest}
	\vspace{-0.3cm}
\end{figure*}

\subsection{Microbenchmarks}
\label{sec:eval_micro}

We identified two main factors impacting the performance overhead, i.e., (i) the number of trusted OS calls and (ii) the complexity/duration of the operation running in the secure world (mainly the logic of the TA). Workloads encompassing simple TA operations/logic with a higher number of world switches will translate into significant performance overhead. In contrast, workloads encompassing complex TA operations with small number of world switches will translate into negligible performance overhead. Multiple zones support has also a negligible impact on the overall performance of the system.

\mypara{World switch time.}
The world switch time increased from an average of 5.7$\pm$0.33\% to 72.5$\pm$1.7\% microseconds, i.e., 13$\times$. This increase is the cumulative penalty of (i) additional \project trampoline and gatekeeper execution time, (ii) cluster-microcontroller requests, (iii) penalty of running a portion of the trampoline with caches disabled, and (iv) time of cache maintenance operations. Despite the high overhead of raw world switches, they often translate into relatively small slowdowns in real-world TEE usage scenarios (see $\S\ref{sec:eval_TAs}$).

\mypara{GlobalPlatform API.}
Figure~\ref{fig:api} presents our main results. The geometric mean of the normalized performance overhead for the nine API benchmarks is 2.327$\times$, i.e., a significant decrease compared to the 13$\times$ overhead from the world switch. We can observe that the three APIs (i.e., \textit{OpenSession}, \textit{InvokeCommand}, and \textit{CloseSession}) that explicitly call the secure world have a larger penalty, ranging from 2.25$\times$ to 4.79$\times$. The \textit{InvokeCommand} and \textit{CloseSession} APIs are the ones that take less time to execute, and thus the ones with the bigger overhead, due to the over-penalty of cache-related operations (see xtest evaluation). Third, from the remaining six APIs, two have almost no penalty. This fact is related to the OP-TEE configuration which disables dynamic shared memory. With this option, \textit{Allocate} and \textit{Register} API implementations are identical. While calling these APIs, there is no switch to the secure world. Therefore, after the execution of \textit{Allocate} and \textit{ReleaseAllocatedSharedMemory}, code and data are already cached for the \textit{Register} and \textit{ReleaseRegisteredSharedMemory} calls. Lastly, we can observe that the system configuration without preemption (RZ No IRQ) slightly decreases the overhead, reducing the geometric mean from 2.327$\times$ to 2.325$\times$.

\mypara{xtest suite.} Figure~\ref{fig:xtest} summarizes our results, represented by the geometric mean of all tests for the same group. Unit tests are represented in nine groups on the left, while benchmarks are represented in three groups on the right. A detailed view of the results per test/benchmark can also be found in the supplementary material published online~\cite{rezonesite}. The group of tests that has a bigger impact is the TEE Internal API and Global Platform API, i.e., 2.97$\times$ and 3.40$\times$, respectively. Note that this group of tests are the ones with smaller execution times and simpler secure world operations. On the other hand, tests such as mbed TLS and key derivation perform fewer but longer operations in the secure world. Since the execution time of the operations performed in the secure world is large, the performance overhead is relatively lower when compared to the TEE API groups. To complete, we tested a potential application-level optimization. Specifically, we modified one particular unit test (4007 symmetric) which consists of performing different symmetric encryption schemes (e.g., AES, DES) for incremental key sizes (e.g., 64-, 128-bit). The standard implementation has a total of 3884 secure world calls per run. We modified the implementation to perform all iterations in a single TA call. By simply batching all operations and processing them in a one-time operation, we reduced the number of calls to the secure world to 96 and were able to decrease the performance overhead from 3.65$\times$ to 1.09$\times$.

\mypara{Multiple zones penalty.} To assess the performance penalty of supporting multiple zones, we built two specific test cases using xtest. The main difference, implementation-wise, to schedule a different zone, is the extra operation required to invalidate the TLB. Thus, in the first experiment, we measured the execution time of running xtest 4001 on trusted OS$_\textrm{2}$, while previously running also trusted OS$_\textrm{2}$. For the second test, we measured the execution time of running xtest 4001 on trusted OS$_\textrm{2}$, while previously running trusted OS$_\textrm{1}$. For the former, it takes, on average, 76.3ms to complete. For the latter, it takes, on average,  78.0ms to complete, i.e., a 1.7ms (2.2\%) increase.

\subsection{Real-world Applications}
\label{sec:eval_TAs}

Our findings suggest that \project will not significantly affect the user experience in real-world applications.

\begin{figure}
    \centering
    \includegraphics[width=0.8\linewidth]{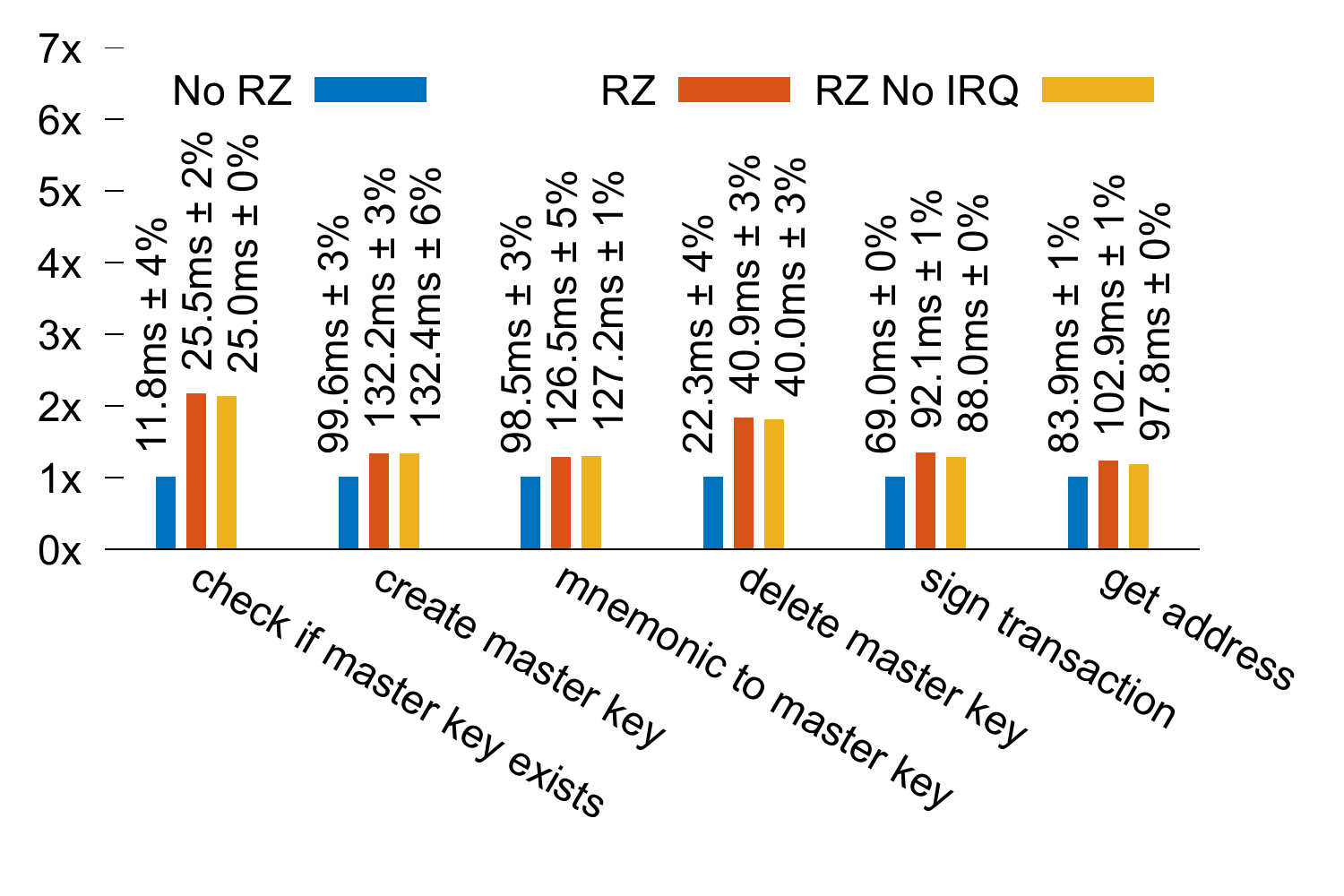}
    \vspace{-.7cm}
    \caption{Relative, absolute execution time, and normalized standard deviation (over 100 runs) for Bitcoin wallet.}
    \label{fig:bitcoinwallet}
	\vspace{-0.3cm}
\end{figure}

In the case of the Bitcoin wallet, the geometric mean of the performance overhead is 1.49$\times$ for the vanilla \project implementation and 1.46$\times$ for a non-preemptible configuration (Figure~\ref{fig:bitcoinwallet}). The most impacted wallet services are related to simple one-time operations. For instance, the absolute execution time of the operation checking if a master key exists is considerably smaller (11.8ms for the standard OP-TEE deployment) than the other services, and thus the penalty is higher. 
Notwithstanding, the results clearly demonstrate that the impact of \project in longer time-consuming operations, e.g., Bitcoin sign transaction (the one likely to be executed more frequently) is low (1.34$\times$) and one order of magnitude smaller than the overhead observed in the raw world switch.

\begin{figure}[t]
    \centering
    \includegraphics[width=0.85\columnwidth]{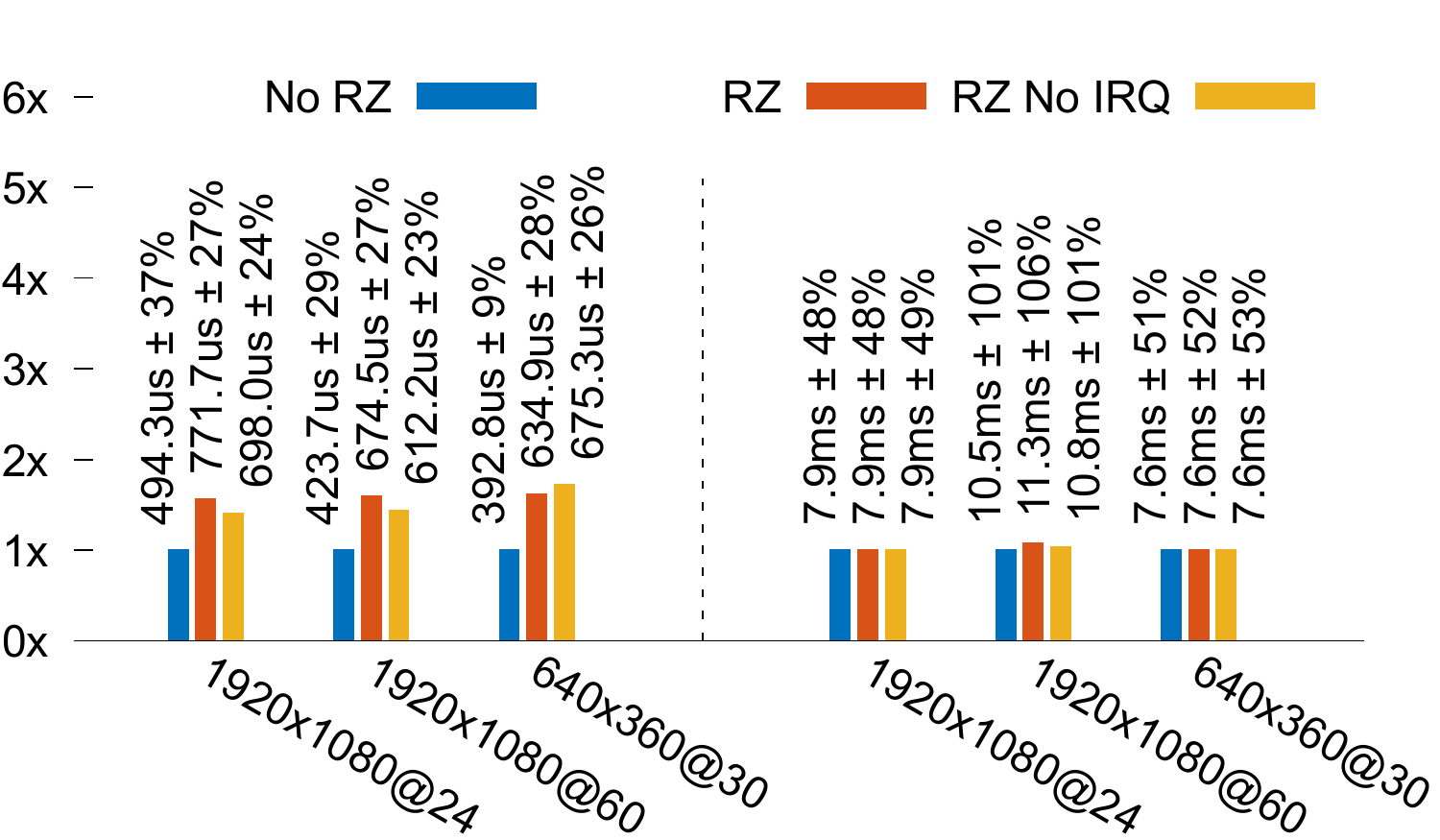}
    \vspace{-0.35cm}
    \caption{Performance overhead, execution time, and normalized standard deviation (over N runs) for DRM subsample decryption (left) and processing (right). N=1000, 1000, 400, for 1080@24, 1080@60, 360@30 setups, respectively. % respectively.
    }
    \label{fig:drm_macro}
    \vspace{-0.3cm}
\end{figure}

Regarding the DRM player, Figure~\ref{fig:drm_macro} plots our results. Overall, the geometric mean of the normalized performance overhead for the DRM subsample decryption is 1.59$\times$. For instance, Figure~\ref{fig:drm_macro} (left) shows that the performance penalty of individual subsample decryption for 1920x1080@60 added by \project (red bar) in comparison to the bare TrustZone setup (blue bar) is 1.59$\times$. This operation is performed every time the secure world is invoked to decrypt a single subsample and return to normal world. These per-subsample overheads tend to be dwarfed when we consider the total processing time that it takes to render a video (Figure~\ref{fig:drm_macro}, right). In this case, computing the time overheads for all configurations gives time increases of 0.2\%, 0.4\%, and 7.7\%, respectively, for 24, 30, and 60 frames per second. The higher the frame, the higher the overhead due to the larger number of trusted OS calls and the smaller duration of secure operations (smaller subsample size).
Second, the \project configuration without preemption (RZ No IRQ) slightly improves the performance of the decryption process, reducing the geometric mean of the normalized performance overhead to 1.52$\times$. This benefit is also noticeable in the processing of the decoded frames and the rendering of the final image (Figure~\ref{fig:drm_macro}, right). Third, although there is a slowdown in decryption, this penalty is in the order of 100 $\mu$seconds. Hence, it is not noticeable in the playback of the video and does not impact the user experience.

\subsection{Impact on the REE Performance}
\label{sec:eval_NW}

We observed that the performance impact from \project on the REE is low, even when stressing secure world calls and the number of parallel calls from concurrent CPUs. We configured the REE to run PCMark (in the quad-core configuration), while concurrently running a toy TA at different intervals (10, 100, and 1000 ms). TA performs a simple addition and returns to normal world. We compared with a setup running vanilla OP-TEE and without calling the TA while running PCMark. Table~\ref{tab:pcmark}, on the left, presents the results. For reasonable intervals between the toy TA calls (1000 and 100), the performance penalty is low, 0.74\% to 4.86\%, respectively. When the interval is 10 ms (similar to a DRM workload), the impact is reasonable (12.97\%) and in the order of magnitude as the ones presented in $\S\ref{sec:eval_TAs}$. To understand the impact on REE for having multiple CPUs issuing concurrent TA calls, we fixed the interval rate of calling the toy TA in 1000ms. Table~\ref{tab:pcmark}, on the right, shows that the impact is low and proportional to the number of concurrent cores, achieving the largest penalty (3.73\%) when the four cores are issuing concurrent TA calls.

\input{TablePcmark}

\mypara{Memory usage.} Given that \project needs to reserve memory for the secure monitor and for allocating zones where trusted OS stacks will be deployed, less memory becomes available to the REE. In our experiments, the secure monitor requires 192KB of memory, and the default OP-TEE configuration that we use for initializing a single zone requires only 32MB. We used this OP-TEE configuration for all our tests, including running the Bitcoin wallet and DRM player (see $\S$\ref{sec:eval_TAs}). Considering that our platform features 4GB of RAM, this means that the memory footprint of \project for hosting a single trusted OS is residual. Naturally, the amount of reserved memory can be increased in order to accommodate i) single trusted OS setups where TAs have higher memory demands, or ii) multiple trusted OS stacks enclosed inside independent zones.

%=======================================================================
%> [SECURITY ANALYSIS]
%=======================================================================

\section{Security Evaluation}
\label{sec:security_analysis}

\subsection{Theoretical Security Analysis}
\label{sec:sec_analysis}

To evaluate the security of \project, we first provide a theoretical assessment of how our system enforces the security properties P1-3 stated in $\S$\ref{sec:threat_model}. We discuss the main attacks that may be attempted to subvert said properties and highlight \project's mechanisms responsible for thwarting them:

\mypara{A1. Direct memory mapping violations.} As described in $\S$\ref{sec:challenge}, an attacker controlling the trusted OS running in S.EL1 of a given zone, may attempt to directly map unauthorized memory regions pertaining to monitor, other zones, or normal world. We prevent these attacks using a PPC to restrict access to memory resources depending on whether a zone is executing (see Figure~\ref{fig:HZ-exec-model}). To protect the normal world, we further need to overcome some limitations of the PPC (see $\S\ref{sec:cross_interference}$) by having only one active core in the cluster while executing a zone, and performing cross-core synchronization (see $\S\ref{sec:synch}$).

%\mypara{A2. Bus master manipulation.} A trusted OS cannot access protected memory. However, an attacker can try to leverage other bus masters to circumvent these protections. To prevent this attack, \project allows a developer to explicitly bind bus masters to zones. This results in a bus master only being able to access resources that are also accessible to the zone.

\mypara{A2. PPC hijacking.} The adversary may attempt to circumvent these restrictions by controlling the PPC and disabling memory protections through reconfiguration of PPC's permissions. To prevent this attack, we ensure that only the trampoline can access the PPC by requiring the trampoline to authenticate itself before using the PPC (see $\S\ref{sec:dynamic-conf}$). This authentication involves a shared secret token established at boot time. A malicious zone can submit a request to unlock the PPC; however, an attacker has a 1 in $2^{64}$ chance of guessing the token correctly. %Appropriate measures can be taken when authentication fails, e.g., returning execution to the monitor.

\mypara{A3. Unintended cache data leakage.} Instead of trying to compromise the PPC through trampoline impersonation, the attacker may leverage cached data to gain access to the secret token. To mitigate this attack, we make sure that the token is only accessible to EL3. This is guaranteed by: i) flushing the token from the caches, ii) having the PPC deny access to the token, and iii) storing the secret token in an exclusive EL3 register while a zone is executing, to allow the trampoline to authenticate itself after a zone exits. The adversary may also leverage cached data to access other memory content from monitor, other zones, or normal world. We perform cache flushes thus evicting sensitive cached data. Memory accesses will then result in direct accesses to main memory, which the PPC can control and block in case of permission violations.

\mypara{A4. Cache code injection.} The attacker can try to bypass the protections described above by tampering with the trampoline at runtime. Although direct access to monitor or gatekeeper memory is not possible, he may try to inject code into the trampoline by leveraging the fact that the PPC does not perform access control at cache level (see $\S\ref{sec:safe_exit}$). To prevent this attack, we disable the MMU for EL3 before entering a zone, thus preventing the trampoline from fetching maliciously manipulated code, and data from being fetched upon a zone exit, allowing the trampoline to sanitize data before continuing.

\mypara{A5. TCB tampering.} Lastly, the adversary may try to disable the protections described above by tampering with \project's TCB, i.e., monitor, gatekeeper, and trampoline, before the system bootstraps, e.g., by replacing a legitimate binary image with another one. Secure boot prevents this attack by aborting the system boot sequence if firmware integrity checks fail.

\subsection{CVE Mitigation Analysis}

Lastly, beyond our theoretical security assessment, we aim to study how deploying \project in real-world systems could help mitigate privilege escalation attacks resulting from the exploitation of vulnerabilities located in TEEs.
As part of this study, we leverage the database of TEE-related CVEs from our previous work~\cite{Cerdeira2020}. We then analyzed 80 CVEs reported as critical and grouped them according to the affected component: trusted OS (TO), trusted application (TA), cryptographic implementation (CI), hardware issue (HW), and bootloader (BL). CVEs that do not clearly identify the vulnerable component (i.e., trusted OS or TA) are labeled as TO/TA. We excluded four CVEs that lack enough information about the affected component. Table~\ref{tab:security-table} summarizes our analysis.

\input{secure_analysis_table}

\input{TableResearchComparisonTCB}

\mypara{Mitigation of attacks in scope.} In total, we identified 66 CVEs that are in scope. These refer to vulnerabilities affecting the trusted OS or trusted applications that have the potential to be successfully exploited and may lead to privilege escalation attacks as indicated in Figure~\ref{fig:intro_trustzone}. In all these cases, \project can help counter successful exploits of these bugs by sandboxing the trusted OS inside a zone and preventing privilege escalation into normal world, secure monitor, or other TAs. In some cases, the attacker may be originally driven by a slightly different goal. For instance, \cve{2018-5210} describes a vulnerability that allows an attacker to gain TEE privilege execution, which leads to retrieving device unlocking information at the TA level. This can then be used to obtain the device unlocking code. We consider this vulnerability sandboxed because an attacker cannot affect other system components. 

\mypara{CVEs out of scope.} In ten cases, CVEs refer to vulnerabilities that fall outside \project's scope, and therefore our system offers limited defenses. 
\cve{2015-9199} only affects specifically shared memory regions. In \cve{2017-18310}, a TA exposes too many services to the normal world, which may allow the normal world to perform abusive actions. \project is not meant to control which operations are available to the normal world. 
\cve{2017-18128} reports a bug in the configuration of a hardware component that can expose secret information. \project is not designed to not prevent the trusted OS from exposing secrets, e.g., to the normal world. In \cve{2017-18282}, a hardware bug allows the normal world to perform secure data accesses; \project does not protect against hardware bugs. Cryptographic-related vulnerabilities such as \cve{2017-14911} are out of scope. \project relies on the correct implementation of cryptographic primitives and schemes. Bootloader-related vulnerabilities, \cve{2016-10458} for example, are also out of scope. \project relies on the platform's secure boot to correctly initialize the system.

%=======================================================================
%> [Comparison with other systems]
%=======================================================================
\section{\project in Perspective}

\subsection{\project and Related Systems}
\label{subsec:related-systems}

In this section, we put \project in perspective with closely related research on TrustZone-aware TEE systems. We surveyed seven representative projects that have been designed to improve the security of TEEs along different (and sometimes complementary) dimensions. Next, we provide an overview of these systems and then compare them with \project according to five distinct criteria. Table~\ref{tab:tcb_comparison} guides our discussion.

\mypara{Related TrustZone-aware TEE systems.} One class of solutions aims to provide normal world enclaves using virtualization. Notable examples include vTZ~\cite{Hua2017}, OSP~\cite{Cho2016}, PrivateZone~\cite{Jang2018} which leverage normal world's virtualization extensions (NS.EL2) to create isolated environments in normal world. While vTZ virtualizes the full TrustZone hardware (enabling the execution of full-fledged trusted OSes in the normal world), OSP and PrivateZone provide a custom runtime environment per TA. Other systems like Sanctuary~\cite{Brasser2019} and TrustICE~\cite{Sun2015} provide normal world enclaves using the TZASC. They rely on monitor and trusted OS to dynamically program the TZASC and create normal world enclaves isolated from the REE. A third class of systems aims to create software-enforced secure world enclaves. For instance, TEEv~\cite{Li2019} and PrOS~\cite{Kwon19} aim to support multiple TEE stacks on Arm platforms where S.EL2 is not available. These solutions rely on a ``hypervisor'' running in S.EL1 (TEEv) or EL3 (PrOS) and perform binary instrumentation of the trusted OSes running in S.EL1 to guarantee isolation. In our work, we introduce a fourth category of systems aimed at creating hardware-enforced secure world enclaves. By relying on PPC/ACU hardware, \project creates secure world enclaves (i.e., zones) to effectively restrict S.EL1 privileges.

\mypara{Security.} To better understand the security properties offered by \project, we consider the threat model defined in $\S\ref{sec:threat_model}$. Assuming that an attacker manages to hijack the trusted OS running in S.EL1, we intend to analyze if the studied systems can prevent further privilege escalation attacks to normal world (P1), secure monitor (P2), or other trusted OSes (P3). For systems that create normal world enclaves, given that the trusted OS hosted by an enclave runs in NS.EL1 (not S.EL1), these attacks are out of scope. Focusing on systems designed to create secure world enclaves, TEEv and PrOS rely on same privilege isolation techniques (binary instrumentation) to shield trusted OS instances running in S.EL1. Although these systems can offer protection against said attacks, they require substantial manual engineering effort or the employment of binary instrumentation tools for removing privileged memory instructions from the trusted OS. In contrast, \project offers similar protections without the need to re-engineer existing trusted OSes, being able to secure unmodified TEE stacks.

\mypara{Scalability.} By scalability we mean to identify important deployability barriers on real-world platforms. We base our analysis on assessing system dependencies on specific hardware components, namely: PPC, ACU, or hardware unavailable on COTS platforms. The fewer these dependencies, the better a solution can scale. Most systems, including \project, depend on some PPC-like component, e.g., SMMU, to restrict accesses from system bus masters. \project also depends on an ACU. Sanctuary precludes PPC or ACU, but relies on the unrealistic assumption that the core ID is propagated to the bus, i.e., this solution is not practical for real-world platforms.

\mypara{Programming overheads.} TrustZone-aware TEE systems may require modifications to standard TEE APIs (e.g., Global Platform) and/or runtime components, i.e., trusted OS and normal world software. Some systems require modifications of trusted OS ~\cite{Li2019, Brasser2019} or normal world hypervisor~\cite{Hua2017}, or implementation of custom drivers~\cite{Jang2018, Cho2016, Brasser2019, Sun2015, Kwon19}. Conversely, \project requires no modifications to TEE APIs or runtime, being able to function with existing TEE/REE stacks.

\mypara{Performance overheads.} Given the heterogeneity of the studied solutions (e.g., target different hardware) and the lack of standardized benchmarks for TrustZone-aware systems, it is difficult to quantitatively compare the performance of these systems. Alternatively, we conducted a qualitative analysis by identifying three types of expensive sources of overhead: TLB misses due to using stage-2 virtualization (S2-TLB), micro-architectural maintenance operations, and trap and emulation. \project does not leverage any hardware virtualization support and does not require trap and emulation. However, it relies on micro-architectural maintenance operations which may sensibly impact the system performance.

\mypara{Trusted computing base.}
\label{sec:eval_tcb}
Different systems depend on various software components to function properly and enforce the security properties for which they were originally designed. In Table~\ref{tab:tcb_comparison}, we identify the components that pertain to the TCB of each system and indicate their respective sizes. We collect this information from the original papers and added reference sizes for TF-A (24,607 SLOC) and OP-TEE (230,094 SLOC). SLOC values were computed using the SLOCCount tool. Modifications to software are marked with ``$^+$''. In \project, the TCB covers a TF-A version that includes the trampoline (1,523 SLOC) and the gatekeeper (3,822 SLOC). \project features a TCB size of about 30 kSLOC that approximates the TCB size of systems that achieve comparable security goals, i.e., TEEv and PrOS. Given that \project's gatekeeper leverages readily available board support package (BSP) code, we foresee that an optimized assembly implementation would reduce gatekeeper's size to just a few hundred SLOC.

\subsection{\project and Armv8.4 S.EL2}
\label{sec:sel2}

So far, we have discussed how \project can reduce the excess of privileges of S.EL1 on Armv8-A platforms prior to the Armv8.4 release. In this specific release, Arm introduced hardware virtualization in the secure world, by extending the hardware architecture with S.EL2, i.e., a new protection mode for hosting a secure hypervisor. Sitting on top of the secure monitor, the secure hypervisor virtualizes the TEE stack by allowing multiple TEE instances to run in isolation. This means that, from releases Armv8.4 onward, S.EL2 seems to provide an alternative solution to implement \project's zones without the need for extra hardware components -- PPC and ACU. By hosting TEE stacks inside independent guest virtual machines (VMs), S.EL2 apparently offers similar security properties as \project's preventing a compromised trusted OS from escaping its VM, therefore protecting normal world (P1), secure monitor (P2), and other TEE instances (P3).

However, despite the introduction of S.EL2, a fundamental violation of the principle of the least privilege still persists: \textit{the secure hypervisor has unlimited privileges to arbitrarily map memory regions into the S.EL2 address space} -- including memory pertaining to monitor and normal world. As a result, if an attacker manages to exploit a bug in the secure hypervisor and run code at S.EL2, Armv8.4 offers no defense mechanism that can further prevent the attacker from violating all properties P1-3 and controlling the entire system. Hijacking the secure hypervisor in such a way is not unrealistic given its relative complexity. For instance, we computed the TCB size of Hafnium~\cite{hafnium}, Arm's reference implementation for the secure hypervisor, and it features over 20 KSLoC. \project can still be leveraged in this setting to isolate the secure hypervisor thus fully guaranteeing P1 and P2. From a technical point of view, repurposing \project should not require major challenges. \project's trampoline can even be simplified as it only needs to shield a single S.EL2 component (akin to a single zone) and interpose context-switch events.

\subsection{\project and Armv9 CCA}
\label{sec:cca}

In Q2 2021, Arm introduced the Confidential Computing Architecture (CCA) whereby Armv9 CPUs are augmented with a so-called Real Management Extension (RME)~\cite{armv9realms}. RME allows ordinary software developers to instantiate a new type of isolation environments named \textit{realms} where they can run application code. To support realms, RME extends the TrustZone architecture with two additional worlds: \textit{root world} and \textit{realm world}. The root world is exclusively dedicated to EL3 and it hosts the secure monitor. The realm world corresponds to a second independent instance of the secure world which is now dedicated to hosting realms. Realm world, secure world, and normal world can run code in EL0, EL1, and EL2 protection modes. In the realm world, EL2 can house a \textit{realm manager} -- equivalent to secure hypervisor -- which in turn can instantiate multiple realms -- i.e., confidential VMs.

Compared to Armv8, CCA introduces important security improvements. For one, the root world prevents access to EL3 memory from any other world. This architectural change effectively solves the excess of privileges of Armv8's S.EL2 or S.EL1 by preventing a compromised secure hypervisor or trusted OS from hijacking the secure monitor, therefore enforcing P2. Similar to Armv8, CCA can also guarantee P3 given that S.EL2 can virtualize S.EL1.

Notwithstanding, \textit{the normal world can arbitrarily be accessed by S.EL2 in secure or realm worlds}. This excess of privileges opens the door for P1 violations due to a compromised realm manager or secure hypervisor. Given CCA's early stage, it is premature to make a definitive assessment of this technology. Nevertheless, our preliminary analysis leads us to infer that \project may be sided with RME to further protect the normal world. Moreover, given that \emph{RME is an optional feature starting only from Armv9.2} \cite{armv9arm}, it is not yet clear how widely OEMs will adopt RME. As of early 2022, the only announced SoCs with Armv9 CPUs feature Armv9.0~\cite{mobilev9}, which omit RME. Without RME, CPUs lack root world protection for EL3~\cite{norealmnoroot} thus having the same security limitations of Armv8.4 CPUs, which may justify deploying \project.

\subsection{Discussion}
\label{sec:discussion}

\mypara{\project limitations and optimizations.} \project, has two central limitations. First (a), since the core ID is not propagated to the bus, only a single TA can simultaneously run per cluster (see $\S\ref{sec:cross_interference}$), which may limit TA concurrency and occasionally increase TA response time. Second (b), frequent invocation of (small duration) TAs may lead to non-negligible performance overheads due to numerous world-switch calls. To improve (a), we propose to enforce a fair-cluster scheduling policy, which balances TA execution across clusters. Since multi-core platforms typically provide multi-cluster configurations (e.g.,
Qualcomm 8 series SoCs since 2016~\cite{qc_multicluster,qc_multicluster2}), this optimization allows concurrently running as many TAs as the available clusters. To improve (b), we suggest batching requests from the rich OS and serving each batch through a single secure world call. This approach reduces the number of world transitions, and consequently the number and impact of \project's microarchitectural maintenance operations.
%\rev{Another limitation is the fact that \project only configures zones at boot time. However, Arm’s reference implementation for the S-EL2 software~\cite{hafnium}, also configures the S-EL1 partitions at boot time. Thus, we don’t see this as a significant limitation. Furthermore, in \project there’s no fundamental limiting factor that prevents zones from being created dynamically, although supporting this feature would add complexity to the secure monitor.}

\mypara{Effects of \project's appropriation of PPC/ACU.} We deduce that \project can leverage a platform's PPC and ACU without jeopardizing the utilization of these components for their original purposes and legacy use cases. In particular, PPC/ACU can be shared, e.g., using a spare security domain in the PPC or enabling the co-existence of the gatekeeper code with other critical or non-critical functionalities already embedded in the ACU. Numerous works in the literature provide strong isolation for software running in microcontrollers \cite{Kwon2019,Pinto2019_2,Hassaan2019,Pinto2020,Grisafi22}. Modern low-end TEEs such as MultiZone~\cite{Pinto2020} leverage minimal hardware primitives ready-available in almost all Arm Cortex-M microcontrollers to provide high-performing, robust, and lightweight enclaves.

\mypara{Coexistence between PPC/ACU and TZASC.} In our current design, \project still leverages the TZASC to protect the secure world from normal world accesses, such as in vanilla TrustZone deployments. Without the TZASC, CPU accesses cannot be differentiated from secure and non-secure. Thus, to protect the normal world, \project would need to perform additional expensive operations (e.g., cross-core synchronization, cluster suspension, cache maintenance operations, etc), with an extra restriction on lack of concurrency with the monitor. A notable exception is TLB maintenance, which would not be necessary given the existence of independent TLB entries for the normal world (that are not shared). In summary, leveraging the TZASC in tandem with the PPC/ACU improves performance while providing security guarantees between normal world and zones.

\mypara{Using Arm SMMU as \project's PPC.}
To leverage an SMMU as PPC, we envision minor changes in the execution flow while entering (EL3$\rightarrow$S.EL1) and exiting (S.EL1$\rightarrow$EL3) a zone. Trampoline's dynamic access control configuration code (see $\S\ref{sec:dynamic-conf}$) must explicitly create page tables that enforce the required access permissions for each zone, instead of configuring permissions through memory-mapped registers. The trampoline must also invalidate the SMMU TLB to ensure the configured page tables enter in effect. The ACU must also prevent the CPU from accessing the SMMU or the page tables otherwise a zone could undo the established access control policy and fully compromise the system.

%=======================================================================
%> [RELATED WORK]
%=======================================================================
\section{Related Work}
\label{sec:related}

%\mypara{Security-oriented hardware architectures.} 
Several security-oriented hardware architectures provide underlying primitives for building TEE systems~\cite{Maene2018, Pinto2019, Cerdeira2020}, namely CPU extensions~\cite{Costan2016_2, Pinto2019, Costan2016}, separate co-processors~\cite{Tarjei2016, SPU2019}, dedicated security chips~\cite{TPM2019}, and secure virtualization~\cite{SEV2019, armv9realms}. Recently, academia has been focused on exploring RISC-V to propose novel TEE hardware architectures~\cite{Nasahl2020, Bahmani2021}. In our work, we leverage COTS hardware primitives to provide augmented isolation within the TEE.

%\mypara{Hardware-enforced TEE systems.}
As for hardware-enforced TEE systems, Komodo~\cite{Ferraiuolo2017} implements a small TEE monitor which provides sealed storage and remote attestation per the SGX specification. 
Lightweight secure world runtimes~\cite{Santos2014,Guan2017} aim at shielding security-critical applications from untrusted OSes. TrustICE~\cite{Sun2015} and Sanctuary~\cite{Brasser2019} leverage the TZASC to create enclaves within the normal world. 
LTZVisor~\cite{ltzvisor} and Voilá~\cite{Pinto2019_2} have leveraged TrustZone hardware to virtualize system resources for a dual OS system configuration while addressing real-time guarantees. \project leverages hardware orthogonal to TrustZone to provide containerization for multiple trusted OSes, while providing the same availability and real-time guarantees as existing TrustZone-assisted TEEs. 
CaSe~\cite{Zhang2016_3} and SecTEE~\cite{Zhao2019} allow TAs to run entirely from the cache and on-chip memory, respectively. Keystone~\cite{Lee2020} and MultiZone~\cite{Garlati2020} leverage standard hardware primitives from the RISC-V ISA, i.e. physical memory protection (PMP), to isolate individual enclaves.  

%\mypara{Virtualization-aided TEE systems.}
vTZ~\cite{Hua2017}, OSP~\cite{Cho2016}, and PrivateZone~\cite{Jang2018} leverage the virtualization extensions available in the normal world (NS.EL2) to create isolated environments. TEEv~\cite{Li2019} and PrOS~\cite{Kwon19} use same privilege isolation \cite{Dautenhahn2015} to secure a minimalist hypervisor from trusted OSes, due to the lack of secure virtualization support prior to Armv8.4-A. Our solution does not rely on any hardware virtualization support and thus can be used in combination with existing hypervisors and legacy systems.

%=======================================================================	
%> [CONC]
%=======================================================================
\section{Conclusion}
\label{sec:conclusion}

With \project, we present a new security architecture that can reduce the privileges of a TrustZone-assisted TEE by leveraging hardware primitives available in modern hardware platforms. We have implemented and evaluated \project for the i.MX 8MQuad EVK and the results demonstrated that the performance of applications such as DRM is not significantly affected. We have also surveyed 80 CVE reports and estimate that \project could help mitigate 86.84\% of them.

%\mypara{\textbf{Acknowledgments.}} We thank our shepherd Aastha Mehta and the anonymous reviewers for their comments and suggestions. This work was supported by national funds through Centro ALGORITMI / Universidade do Minho, Instituto Superior Técnico / Universidade de Lisboa, and FCT under project UIDB/50021/2020 and UIDB/00319/2020. 
% and grant 2020.05270.BD, and via project COSMOS (via the OE with ref. PTDC/EEI-COM/29271/2017 and via the ``Programa Operacional Regional de Lisboa na sua componente FEDER'' with ref. Lisboa-01-0145-FEDER-029271). 
%David Cerdeira was supported by FCT grant SFRH/BD/146231/2019.

%to restrict the privileges of the trusted OS
%creating sandboxed domains (\textit{zones}) inside the secure world. The key idea behind the \project design relies on

% %-------------------------------------------------------------------------------
\section*{Acknowledgments}
% %-------------------------------------------------------------------------------
% 
We thank our shepherd Aastha Mehta and the anonymous reviewers for their comments and suggestions. This work was supported by national funds through Centro ALGORITMI / Universidade do Minho, Instituto Superior Técnico / Universidade de Lisboa, and FCT under project UIDB/50021/2020 and UIDB/00319/2020. 
% and grant 2020.05270.BD, and via project COSMOS (via the OE with ref. PTDC/EEI-COM/29271/2017 and via the ``Programa Operacional Regional de Lisboa na sua componente FEDER'' with ref. Lisboa-01-0145-FEDER-029271). 
David Cerdeira was supported by FCT grant SFRH/BD/146231/2019.

% 
% %-------------------------------------------------------------------------------
% \section*{Availability}
% %-------------------------------------------------------------------------------
% 
% USENIX program committees give extra points to submissions that are backed by artifacts that are publicly available. If you made your code or data available, it's worth mentioning this fact in a dedicated section.
% 
% %-------------------------------------------------------------------------------
%\newpage
{%\footnotesize
\small
\bibliographystyle{plain}
\bibliography{references.bib}
}

%\newpage
%\section*{Appendix}

%Please visit \url{link.com} \david{TODO:} for a fine-grain view of the results collected from the xtest suite. 

\if10
%... detail the overhead for each xtest test we execute to evaluate \project. Again, we advise the reader to not take this results at face value, as only the benchmark tests are intended to serve as benchmark. However the tests are useful in that they provide a diverse range of behaviors and workloads, and by analyzing them, they shed light on the impact of \project on various scenarios.

\label{sec:appdx}

\begin{figure}[!h]
    \centering
    \includegraphics[scale=0.4]{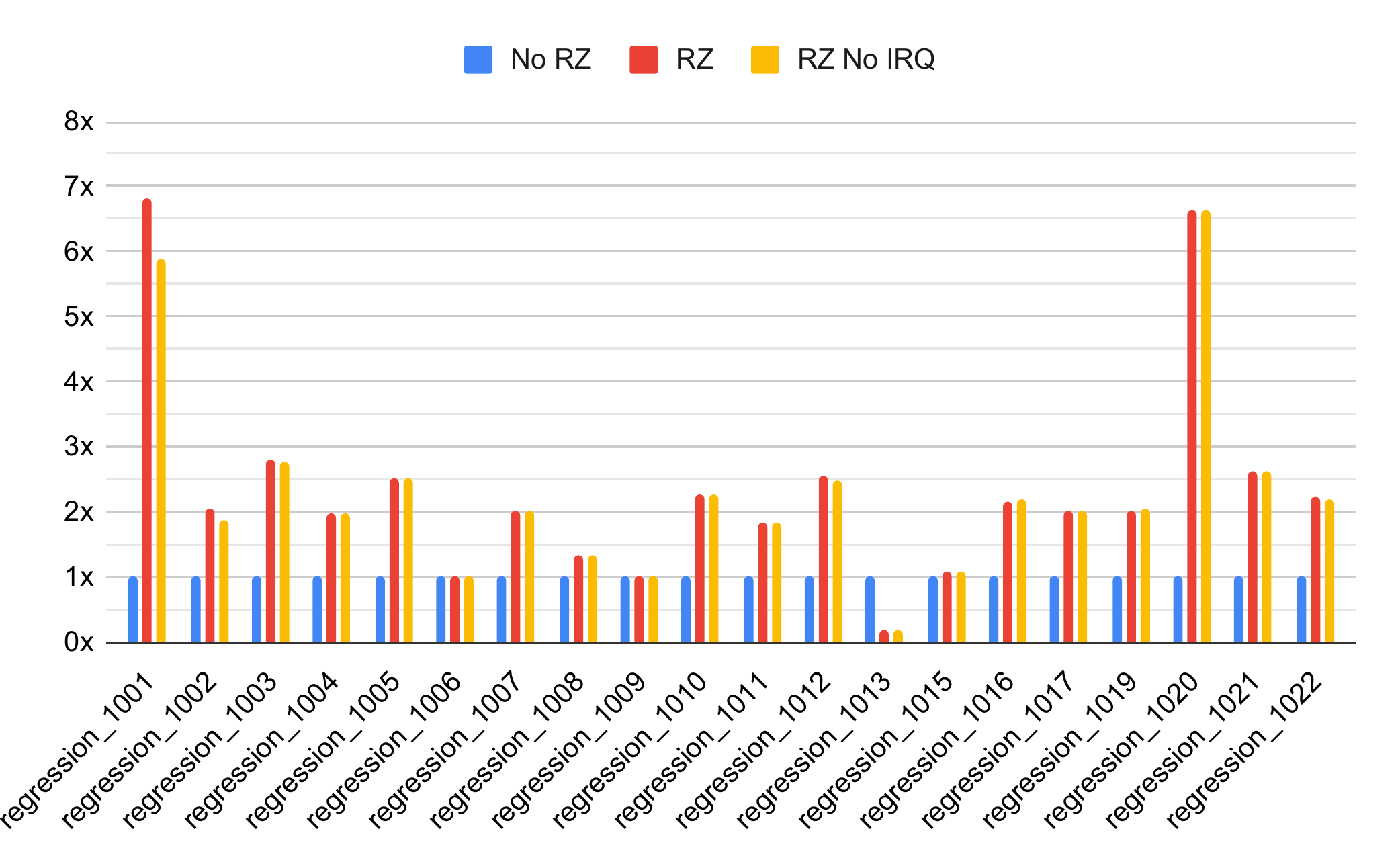}
    \caption{OS core features xtest tests}
    \label{img:appdx_1}
\end{figure}

\begin{figure}[!h]
    \centering
    \includegraphics[scale=0.4]{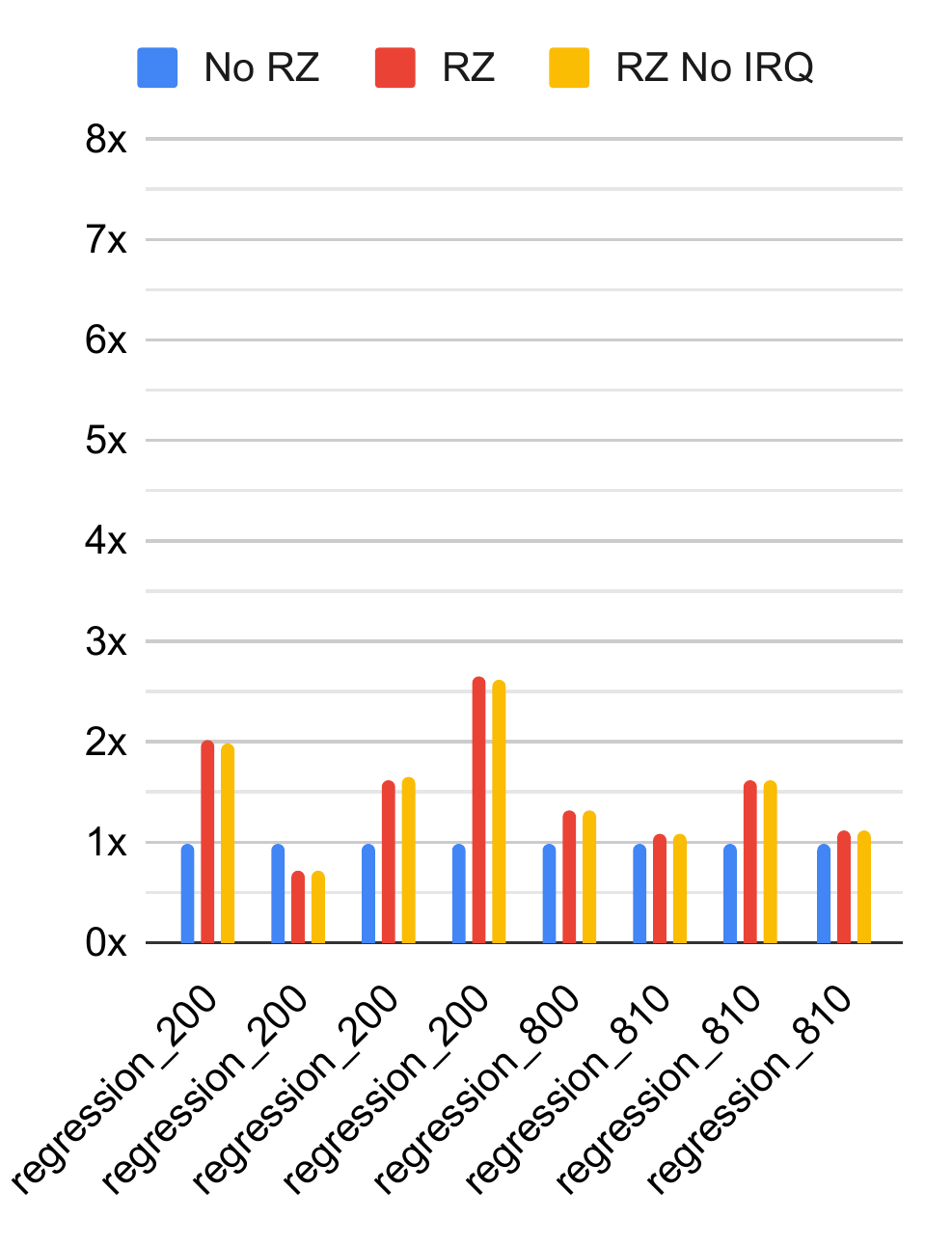}
    \caption{Network socket, key derivation and mbed TLS xtest tests}
     \label{img:appdx_2}
\end{figure}

\begin{figure}[!h]
    \centering
    \includegraphics[scale=0.4]{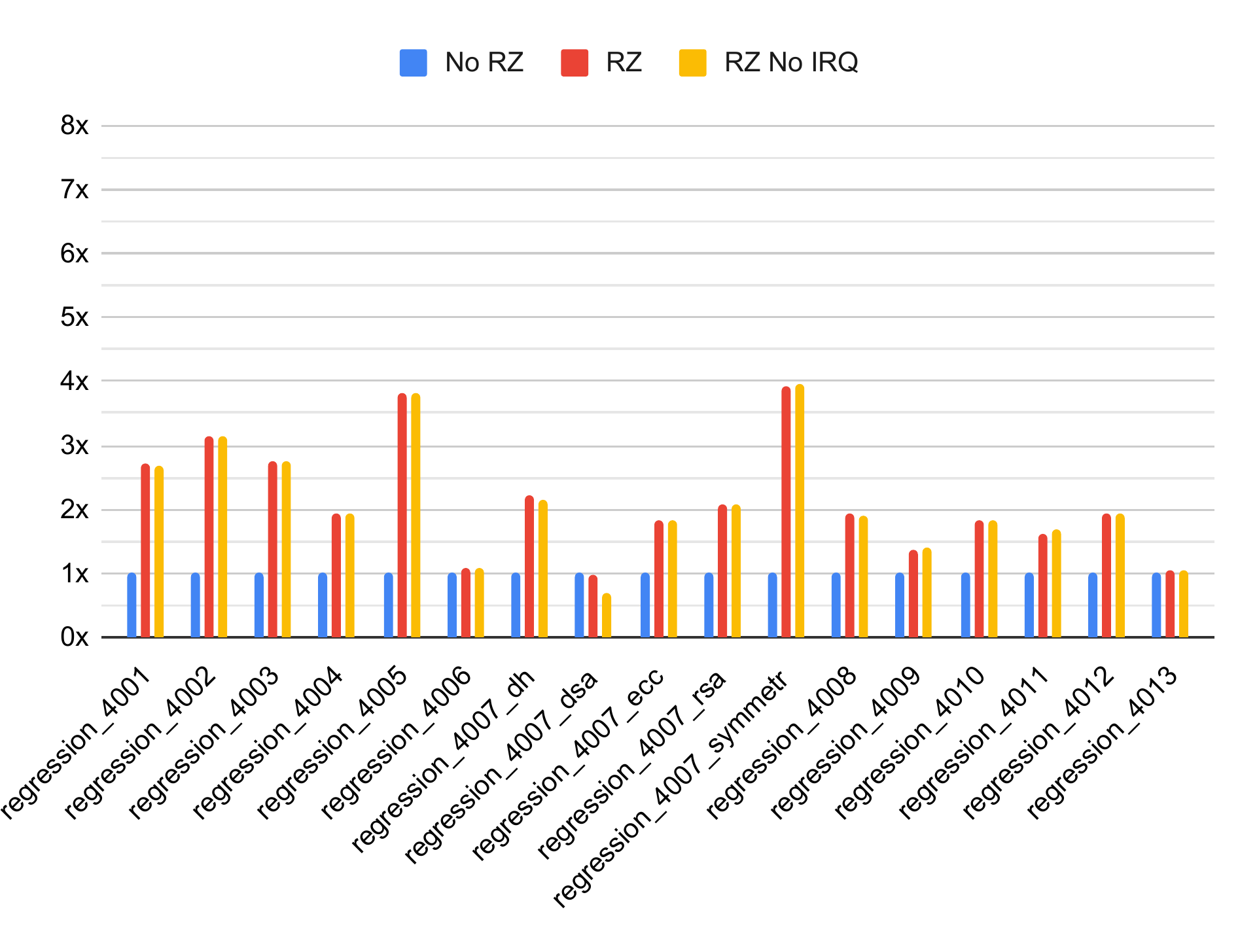}
    \caption{Cryptographic algorithm xtest tests}
     \label{img:appdx_3}
\end{figure}

\begin{figure}[!h]
    \centering
    \includegraphics[scale=0.4]{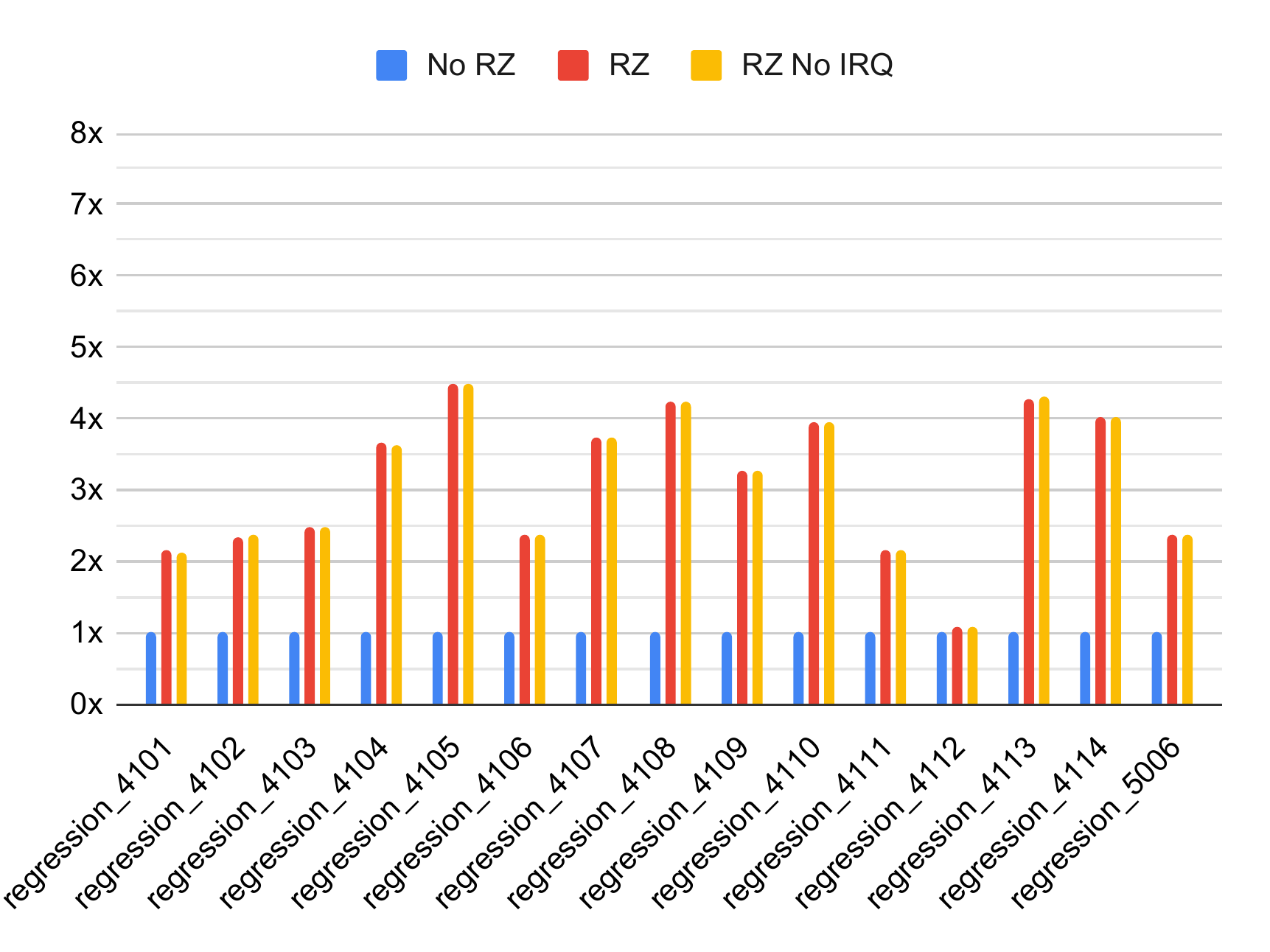}
    \caption{TEE Internal Arithmetic API, and Global Platform API xtest tests}
     \label{img:appdx_4}
\end{figure}

\begin{figure}[!h]
    \centering
    \includegraphics[scale=0.4]{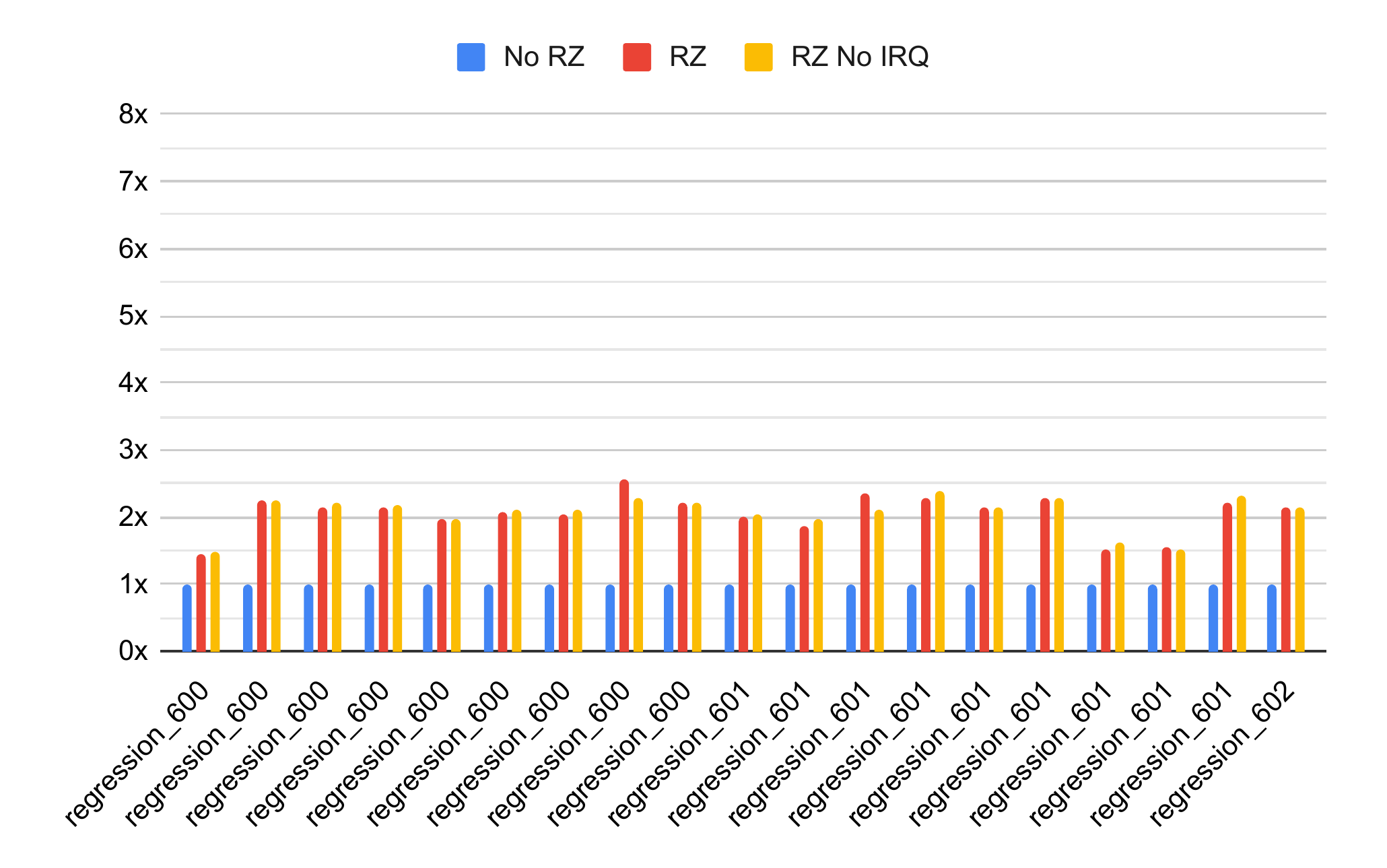}
    \caption{Storage xtest tests}
     \label{img:appdx_5}
\end{figure}

\begin{figure}[!h]
    \centering
    \includegraphics[scale=0.4]{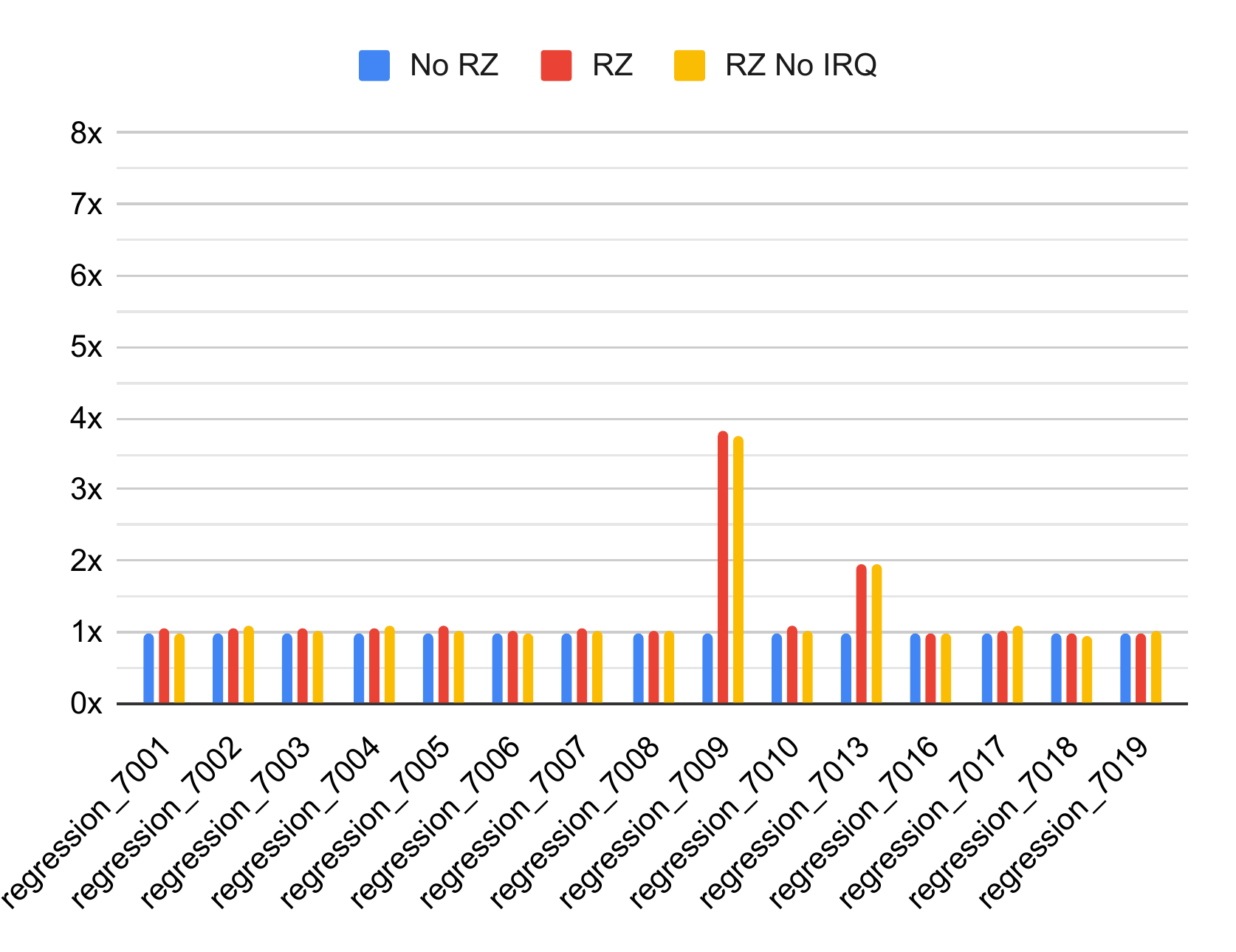}
    \caption{Shared memory xtest tests}
     \label{img:appdx_6}
\end{figure}

\begin{figure}[!h]
    \centering
    \includegraphics[scale=0.4]{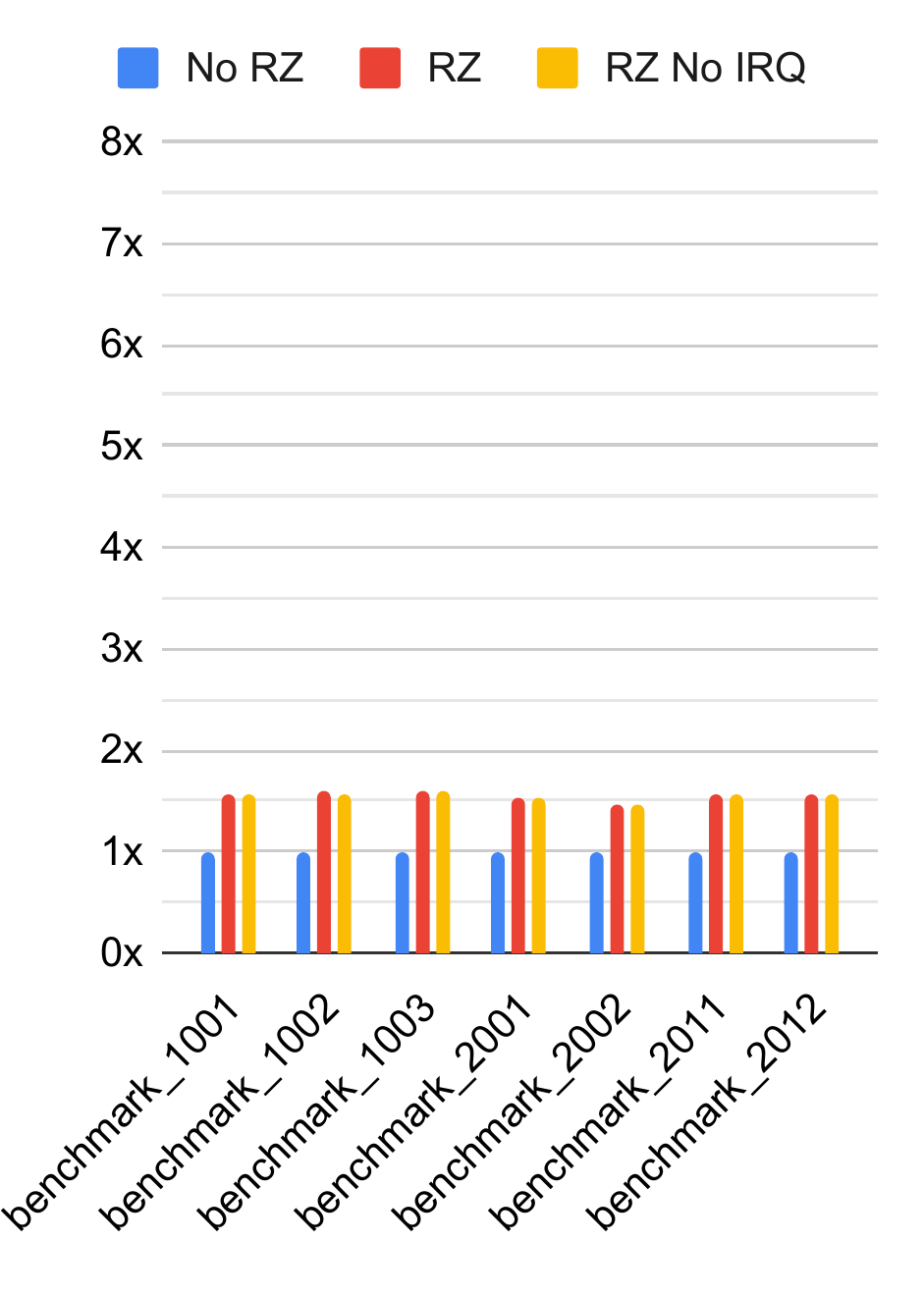}
    \caption{xtest benchmark suite}
     \label{img:appdx_7}
%\vspace{16.3cm}
\end{figure}
\fi

%% file: TableCOTS.tex
\begin{table*}[t]\centering
%\scriptsize
\newcommand{\Y}{\checkmark}
\newcommand{\Par}{\LEFTcircle}
\newcommand{\N}{---}
\newcommand{\II}{???}
\newcommand{\NA}{N/A}
\scalebox{0.75}{
\color{black}
\setlength{\tabcolsep}{3pt}
\renewcommand{\arraystretch}{1.3}
\resizebox{1.3\textwidth}{!}{%
\begin{tabularx}{1.3\textwidth}{p{1.8cm}p{3.6cm}p{2.6cm}Xp{1.2cm}p{11cm}}
\toprule
\textbf{Vendor} & \textbf{SoC Platform} & \textbf{PPC} & \textbf{ACU} & \textbf{RZ?} & \textbf{Sources} \\ \midrule

NXP & iMX8MQ & RDC & Cortex-M4 & Yes & {\small Ref. manual (\url{https://www.nxp.com/webapp/Download?colCode=IMX8MDQLQRM})} \\
& iMX8QM & xRDC & SCU & Yes & {\small Ref. manual (\url{https://www.nxp.com/webapp/Download?colCode=IMX8QMRM})} \\ \midrule

Xilinx & UltraScale+ MPSoC & XMPU, XPPU & PMU & Yes & {\small Ref. manual (\url{https://www.xilinx.com/support/documentation/user\_guides/ug1085-zynq-ultrascale-trm.pdf})} \\  \midrule

Nvidia & Tegra X1 and X2, Xavier & SMMU & BPMP & Yes & {\small Ref. manual (\url{https://developer.nvidia.com/embedded/downloads})} \\ \midrule

Socionext & SC2A11 & SMMU & SCP & Yes & {\small TF-A V2.6 (\url{https://github.com/ARM-software/arm-trusted-firmware}) and U-boot  v2022.01 (\url{https://github.com/u-boot/u-boot})} \\ \midrule

Qualcomm & Snapdragon 845, 855, 865, 888, 8gen1 & SMMU, XPU* & SPU & Yes & {\small Linux 5.16 for SD855 (\url{https://gitlab.com/sm8150-mainline/linux}), \href{https://nvd.nist.gov/vuln/detail/CVE-2020-11252}{CVE-2020-11252}, \href{https://nvd.nist.gov/vuln/detail/CVE-2017-18311}{CVE-2017-18311}, \cite{qualcommxpu}. *We lack the data to support that 845 and 8gen1 contain an XPU, though earlier platforms of the same segment feature one.} \\ \midrule

Broadcom & Stingray & SMMU & SCP & Yes & {\small TF-A (Same as Socionext), Linux 5.16 (\url{https://github.com/torvalds/linux})}\\ \midrule

Samsung & Exynos 990 & Periph. MMUs & iSE & N/A & \multirow{2}{*}{\small S20 and S21 Linux kernel source code (\url{https://opensource.samsung.com/main})} \\ 
& Exynos 2100 & Periph. MMUs & eSE & N/A & \\  \midrule

Mediatek & Dimensity 1200 & \xmark & \xmark & No & {\small OnePlus Nord 2 Linux kernel source (\url{https://github.com/OnePlusOSS/android\_kernel\_oneplus\_mt6893.git})} \\  \midrule

\multirow{1}{*}{HiSilicon} & Kirin 9000 & \xmark & PMU & No & {\small Huawei Mate 40 Pro Linux (\url{https://consumer.huawei.com/en/opensource/})} \\
\bottomrule
\end{tabularx}
}
}
\vspace{-0.1cm}
\caption{Availability of PPC/ACU hardware primitives and applicability of \project on COTS platforms.
} \label{tab:rezone-cots}
\vspace{-0.1cm}
\end{table*}

%% file: TablePcmark.tex
\begin{table}[!t]
\setlength{\tabcolsep}{6pt}
\renewcommand{\arraystretch}{1.1}
\centering
\footnotesize
\scalebox{0.95}{

\begin{tabular}{cc}

\begin{tabular}{ccc}
Interval (ms) & Score & Penalty \\
\toprule
10 & 4369 & 12.97\% \\
100 & 4776 & 4.86\% \\
1000 & 4983 & 0.74\% \\
\bottomrule
\end{tabular} 

&

\begin{tabular}{ccc}
\# of Cores & Score & Penalty \\
\toprule
1 & 4983 & 0.74\% \\
2 & 4938 & 1.63\% \\
4 & 4833 & 3.73\% \\ 
\bottomrule
\end{tabular}  

\end{tabular}

}
\caption{Impact of \project on the PCMark performance score assessing the REE. On the left, the impact of decreasing the interval between calls for one core. On the right, the effect of issuing TA calls (1000 ms intervals) with multiple cores.}
\label{tab:pcmark}
\vspace{-0.2cm}
\end{table}

%% file: secure_analysis_table.tex
%Please add the following packages \II necessary:
%\usepackage{booktabs, multirow} % for borders and merged ranges
%\usepackage{soul}% for underlines
%\usepackage[table]{xcolor} % for cell colors
%\usepackage{changepage,threeparttable} % for wide tables
%If the table is too wide, replace \begin{table}[!htp]...\end{table} with
%\begin{adjustwidth}{-2.5 cm}{-2.5 cm}\centering\begin{threeparttable}[!htb]...\end{threeparttable}\end{adjustwidth}
\begin{table}[!t]
\centering
%\scriptsize
\newcommand{\Y}{\checkmark}
\newcommand{\Par}{\LEFTcircle}
\newcommand{\N}{---}
\newcommand{\II}{?}
\newcommand{\NA}{N/A}
\scalebox{0.61}{
\setlength{\tabcolsep}{3pt}

\begin{tabular}{|ccc|ccc|ccc|ccc|}
\hline
CVE & C & & CVE & C & & CVE & C & & CVE & C & \\
\hline
\cveshort{2014-9979} & TO & \Y & \cveshort{2017-11011} & TO & \Y & \cveshort{2015-9000} & TA & \Y & \cveshort{2015-8997} & TO/TA & \Y \\
\cveshort{2015-8999} & TO & \Y & \cveshort{2017-14912} & TO & \Y & \cveshort{2015-9002} & TA & \Y & \cveshort{2015-8998} & TO/TA & \Y \\
\cveshort{2015-9070} & TO & \Y & \cveshort{2017-14916} & TO & \Y & \cveshort{2015-9162} & TA & \Y & \cveshort{2015-9005} & TO/TA & \Y \\
\cveshort{2015-9071} & TO & \Y & \cveshort{2017-14917} & TO & \Y & \cveshort{2015-9174} & TA & \Y & \cveshort{2015-9007} & TO/TA & \Y \\
\cveshort{2015-9072} & TO & \Y & \cveshort{2017-17176} & TO & \Y & \cveshort{2015-9183} & TA & \Y & \cveshort{2016-2432} & TO/TA & \Y \\
\cveshort{2015-9073} & TO & \Y & \cveshort{2017-18071} & TO & \Y & \cveshort{2017-6293} & TA & \Y & \cveshort{2016-10297} & TO/TA & \Y \\
\cveshort{2015-9108} & TO & \Y & \cveshort{2017-18128} & TO & \N & \cveshort{2017-18310} & TA & \N & \cveshort{2017-6289} & TO/TA & \Y \\
\cveshort{2015-9112} & TO & \Y & \cveshort{2017-18129} & TO & \Y & \cveshort{2017-18312} & TA & \II & \cveshort{2017-14913} & TO/TA & \Y \\
\cveshort{2015-9113} & TO & \Y & \cveshort{2017-18132} & TO & \Y & \cveshort{2017-18317} & TA & \II & \cveshort{2017-18293} & TO/TA & \Y \\
\cveshort{2015-9198} & TO & \Y & \cveshort{2017-18133} & TO & \Y & \cveshort{2018-5210} & TA & \Y & \cveshort{2017-18296} & TO/TA & \II \\
\cveshort{2015-9199} & TO & \N & \cveshort{2017-18311} & TO & \Y & \cveshort{2018-5885} & TA & \Y & \cveshort{2017-18297} & TO/TA & \Y \\
\cveshort{2015-9200} & TO & \Y & \cveshort{2017-18314} & TO & \II & \cveshort{2014-9932} & TO/TA & \Y & \cveshort{2017-18298} & TO/TA & \Y \\
\cveshort{2016-2431} & TO & \Y & \cveshort{2017-18315} & TO & \Y & \cveshort{2014-9935} & TO/TA & \Y & \cveshort{2018-5866} & TO/TA & \Y \\
\cveshort{2016-10238} & TO & \Y & \cveshort{2018-3588} & TO & \Y & \cveshort{2014-9936} & TO/TA & \Y & \cveshort{2017-18282} & HW & \N \\
\cveshort{2016-10432} & TO & \Y & \cveshort{2018-5870} & TO & \Y & \cveshort{2014-9937} & TO/TA & \Y & \cveshort{2015-9003} & CI & \N \\
\cveshort{2017-6290} & TO & \Y & \cveshort{2016-10239} & TO & \Y & \cveshort{2014-9945} & TO/TA & \Y & \cveshort{2016-10398} & CI & \N \\
\cveshort{2017-6292} & TO & \Y & \cveshort{2018-11950} & TO & \Y & \cveshort{2014-9948} & TO/TA & \Y & \cveshort{2017-14907} & CI & \N \\
\cveshort{2017-6294} & TO & \Y & \cveshort{2015-4422} & TA & \Y & \cveshort{2014-9949} & TO/TA & \Y & \cveshort{2017-18146} & CI & \N \\
\cveshort{2017-8274} & TO & \Y & \cveshort{2015-6639} & TA & \Y & \cveshort{2015-8995} & TO/TA & \Y & \cveshort{2016-10458} & BL & \N \\
\cveshort{2017-11010} & TO & \Y & \cveshort{2015-6647} & TA & \Y & \cveshort{2015-8996} & TO/TA & \Y & \cveshort{2017-14911} & BL & \N \\ 
\hline
\end{tabular} 
}

\vspace{0.3cm}

\renewcommand{\arraystretch}{1.1}
\scalebox{0.78}{
\begin{tabular}{ccccccc|cc}
In Scope? & TO/TA & TO & TA & HW & CI & BL & Total & Percentage \\
\hline
Y & 21  & 34    & 11    &   &   &   & 66 & 86.84\% \\
N &     & 2     & 1     & 1 & 4 & 2 & 10 & 13.16\% \\
\hline
Total & 21 & 36 & 12 & 1 & 4 & 2 & 76 & \\ 
\hline
\end{tabular} 
}
%\vspace{-0.1cm}
\caption{Mitigation analysis of selected CVEs with critical CVSS scores: CVEs in scope (\Y), out of scope (\N), insufficient information (\II).} \label{tab:security-table}
\vspace{-0.3cm}
\end{table}

%% file: TableResearchComparisonTCB.tex
\begin{table*}[t]\centering
%\scriptsize
% \setlength{\tabcolsep}{2pt}
\scalebox{0.65}{
\setlength{\tabcolsep}{3pt}
\renewcommand{\arraystretch}{1.2}
\begin{tabular}{cc|ccc|ccc|cc|ccc|cccc}
& & \multicolumn{3}{c|}{\makecell{Security \\ (Can defend against)}} & \multicolumn{3}{c|}{ \makecell{Scalability \\ (Does not depend on)}} & \multicolumn{2}{c|}{\makecell{Programming \\ (Unmod. code)}} & \multicolumn{3}{c|}{\makecell{Performance \\ (Sources of overhead)}} & \multicolumn{4}{c}{\makecell{TCB \\ (Subcomponents and total size)}} \\
\midrule
\makecell{Enclave \\ Type} & System & \makecell{TOS→NW \\ (P1)}
& \makecell{ TOS→Mon \\ (P2)}
& \makecell{ TOS$_1$→TOS$_2$ \\ (P3)}
& PPC & ACU & \makecell{Non-COTS \\ Features} & Runtime & API & S2-TLB & \makecell{Micro. Arch. \\ Maintanance} & \makecell{Trap \& \\ Emul.} & Monitor & Trusted OS & \makecell{Other \\ SLOC} & \makecell{Total \\ kSLOC} \\
\midrule

%& Typical TZ TEE & --- & --- & --- & \good & \good & \good & \good & \good & \good & \good & \good & TF-A & OPTEE & - & 254.7 \\
\multirow{5}{*}{ NW } & PrivateZone \cite{Jang2018} & --- & --- & --- & \bad & \good & \good & \bad & \bad & \bad & \bad & \good & TF-A$^+$ & OPTEE & - & 259.9 \\
& OSP \cite{Cho2016} & --- & --- & --- & \bad & \good & \good & \bad & \bad & \bad & \bad & \good & TF-A$^+$ & OPTEE & OSP Hyp & 255.5 \\
& vTZ \cite{Hua2017} & --- & --- & --- & \bad & \good & \good & \good & \good & \bad & \good & \bad & Custom & - & - & 2.0 \\
& Sanctuary \cite{Brasser2019} & --- & --- & --- & \good & \good & \bad & \bad & \bad & \good & \bad & \good & TF-A$^+$ & OPTEE$^+$ & TA 2x & 256.0 \\
& TrustICE \cite{Sun2015} & --- & --- & --- & \good & \good & \good & \bad & \bad & \good & \bad & \good & TF-A$^+$ & OPTEE & - & 255.0 \\
\midrule
\multirow{3}{*}{ SW } & TEEv \cite{Li2019} & \half & \half & \half & \good & \good & \good & \bad & \good & \good & \good & \bad & TF-A$^+$ & - & TEE-Visor & 28.8 \\
& PrOS \cite{Kwon19} & \half & \half & \half & \bad & \good & \good & \bad & \good & \good & \good & \bad & TF-A$^+$ & - & PVisor & 27.6 \\
\cmidrule{2-17}
& \project & \good & \good & \good & \bad & \bad & \good & \good & \good & \good & \bad & \good & TF-A$^+$ & - & gatekeeper & 30.0 \\
\bottomrule
\end{tabular} 
}
\vspace{-0.1cm}
\caption{Analysis of \project and similar systems: \good\xspace indicates that the system fares positively, \bad\xspace that it fares negatively.}
 \vspace{-0.2cm}
\label{tab:tcb_comparison}
\end{table*}